\documentclass[12pt]{article}
\usepackage[cp866]{inputenc}
\usepackage{textcomp}
\usepackage{amsfonts,amssymb,amsmath,fullpage,pst-node,rotating,graphics,epsfig, subfigure}
\usepackage{multirow}
\usepackage{authblk}

\usepackage{xcolor}
\definecolor{darkgreen}{rgb}{0.,0.3,0.}

\newcommand{\ri}{{\rm i}}

\newcommand{\rM}{{\rm M}}
\newcommand{\ro}{{\rm o}}
\newcommand{\re}{{\rm e}}
\newcommand{\rd}{\,{\rm d}\,}

\newcommand{\cH}{{\cal H}}
\newcommand{\cG}{{\cal G}}
\newcommand{\cF}{{\cal F}}
\newcommand{\bL}{{\bf L}}
\newcommand{\bR}{{\bf R}}
\newcommand{\bs}{{\bf s}}
\newcommand{\rE}{{\rm E}}
\newcommand{\rJ}{{\rm J}}
\newcommand{\rS}{{\rm S}}
\newcommand{\bom}{\mbox{\boldmath${\omega}$}}

\newtheorem{remark}{Remark}

\begin{document}
\title{Impact of a moon on the evolution of\\
a planet's obliquity: a non-resonant case}

\author[1]{O.M.~Podvigina}
\author[2]{P.S.~Krasilnikov}
\affil[1]{\small Institute of Earthquake Prediction Theory
and Mathematical Geophysics, Russian Academy of Sciences,
84/32 Profsoyuznaya St, 117997 Moscow, Russian Federation}
\affil[2]{\small Moscow Aviation Institute , 4 Volokolamskoe shosse,
125993 Moscow, Russian Federation}

\maketitle

\begin{abstract}

We investigate how the variation of the obliquity
(the axial tilt) of a hypothetical exo-Earth is effected by the presence of
a satellite, an exo-Moon.
Namely, we study analytically and numerically how the range of obliquity of the
exo-Earth changes if an exo-Moon is added to a system comprised of exo-Sun,
exo-Earth and exo-planets. We say that the impact of the exo-Moon is
{\it stabilising} if upon the addition of the exo-Moon the range of obliquity
decreases, while we call the impact {\it destabilising}
if the range increases as the exo-Moon is added to the system.
The problem is considered in a general
setup. The exo-Earth is assumed to be rigid, axially symmetric and almost
spherical, the difference between the largest and the smallest principal
moments of inertia being a small parameter of the problem. Assuming the orbits
of the celestial bodies to be quasiperiodic, we apply time averaging to study
rotation of the exo-Earth at times large compared to the respective periods.
Non-resonant frequencies are assumed.
We identify a class of systems for which we prove analytically that the impact
of the exo-Moon is stabilising and a class where it is destabilising.
We also investigate numerically how the impact of the exo-Moon
in a particular system comprised of a star and two planets varies on modifying
the geometry of the orbits of the exo-Moon and the second planet and
the initial obliquity.

\noindent
\bigskip{\bf Key words:} obliquity, exoplanet, averaging, Hamiltonian dynamics
\end{abstract}

\section{Introduction}\label{sec1}

Obliquity, i.e., the orientation of the rotation axis of a planet relative to
its orbit, is an important factor that determines whether a planet is
hospitable to life \cite{ar14,co12,fe14,he11,ki17,sp09}.
Both extremely high and low values of obliquity yield contrast distribution of
temperatures over the globe, while moderate values of obliquity cause seasons,
leading to a more uniform distribution and relatively stable climate.
The obliquity of the Earth varies just between 22.1$^\circ$ and 24.5$^\circ$ and
its orbital eccentricity between 0 and 0.06, but even such small variations
result on the occurrence of glacial/interglacial cycles accompanied by
substantial changes of average temperatures \cite{ka03,mi41,wi03}.
Numerical simulations of Lascar {\it et al} of the evolution of the Earth's
obliquity with and without the Moon \cite{la93,la93b,la93c} revealed that the Moon has
a stabilising effect. If the presence of a heavy moon stabilises the
obliquity of a planet in a general case as well, then the presence of the
moon is strongly beneficial for an advanced life to develop.
Given a large number of exo-planets discovered to the day, presence of
heavy moons may be useful pointers for identification of possibly habitable
planets.
No compelling evidence has been found for exomoons around the observed
exoplanets \cite{ki13a,ki13b}.
Therefore, if a planet is life-bearing only when it is accompanied by a large
moon, this requirement significantly decreases the chances for intelligent life
to develop.

Most of the studies of the influence of a satellite on the evolution of
a planet's obliquity have focused on a particular case of the Earth-Moon
system.
By numerical integration of the equations of precession
it was found that for the moonless
Earth's obliquity would vary chaotically from 0 to 85 degrees
\cite{la93,la93b,la93c,sl97} while in the presence of
the Moon the window of initial values of obliquity resulting on chaotic
behaviour decreases to between 60 to 90 degrees and outside this window the
variations of obliquity are much smaller.
A different result for
the moonless Earth was obtained in \cite{lbc11}, where it was found
that the difference between maximal and minimal values of
obliquity does not exceed 10$^\circ$, leading to the conclusion that
``A large moon thus does not seem to be needed to stabilize the obliquity
of an Earth-like planet on timescales relevant to the development
of advanced life''.
This conjecture was supported by the findings of \cite{lb14},
where analytical estimates of the characteristic Lyapunov exponents
and the chaotic diffusion rate were obtained, ``the stochastic change
in Earth's obliquity is sufficiently slow to not preclude long-time
habitability''.

\begin{figure}[h]
\vspace*{-12mm}
\hspace*{2mm}\includegraphics[width=12cm]{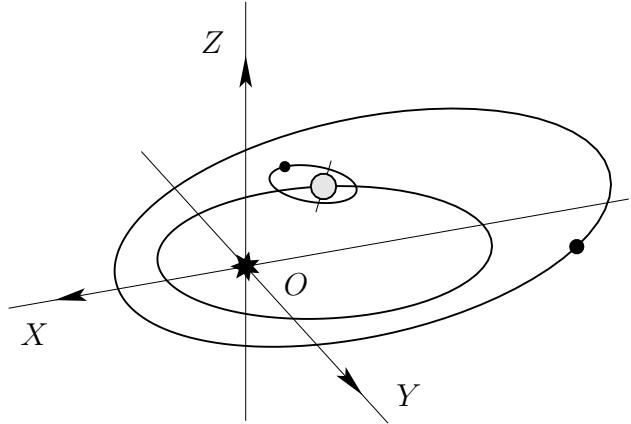}

\vspace*{-55mm}\hspace*{37mm}$X$

\vspace*{3mm}\hspace*{87mm}$Y$

\vspace*{-52mm}\hspace*{61mm}$Z$

\vspace*{27mm}\hspace*{72mm}$O$

\vspace*{20mm}
\noindent
\caption{The planetary system studied in sections \ref{sec5}:
exo-Sun, exo-Earth, exo-Moon and exo-Jupiter.}
\label{fig1}\end{figure}

The evolution of obliquity of a planet not taking into account the influence
of a moon was studied in a large number of papers, e.g., in
\cite{cl03,cls03,la04,kr18,wh04,hw04} for Solar system planets
or in \cite{qu20} for exo-planets.
In such systems the obliquity is often chaotic and/or undergoes
substantial variations in the course of temporal evolution thus supporting the
conjecture about stabilising influence of a moon.
However,
the number of studies of planet with a moon, rather then Earth, is very limited.
In \cite{ai07} obliquity of an exo-planet with a satellite in a one-planet
system was studied taking into account tidal effects. It was found that
for certain values of initial obliquity it can possibly oscillate with large
amplitudes.

In this paper we investigate the influence of a heavy satellite on the variation
of obliquity of a planet in a general setup.
We study numerically and analytically the behaviour of obliquity on large time
scales in a planetary system comprised of a star, planets and a satellite,
orbiting one of them.
This planet and its satellite are called
exo-Earth and exo-Moon, respectively. The exo-Earth is an axially symmetric
rigid body and the difference between the largest and smallest principal moments
is a small parameter. Other celestial bodies are assumed to be point masses.
The planets move along quasi-periodic orbits with prescribed frequencies
$(\omega_1,...,\omega_K)$.
The orbit of the exo-Moon keeps a constant inclination to the ecliptic
and undergoes two types of slow precessional motion,
nodal and apsidal, with respective frequencies $\sigma_n$ and $\sigma_a$.
(The nodal precession is the precession of the exo-Moon's orbital plane and
the apsidal one is the rotation of the exo-Moon's orbit within the plane.)
The frequencies $\bom=(\omega,\omega_1,...,\omega_K)$, where $\omega$ is
the frequency of the exo-Earth rotation, are order one and non-resonant.

In order to define whether the influence of the exo-Moon on the rotation of
the exo-Earth is stabilising or not, we compare the range of obliquity
\begin{equation}\label{Delta}
\Delta(I_0,h_0)=
\sup_{-\infty<t<\infty}I(t,I_0,h_0)-\inf_{-\infty<t<\infty}I(t,I_0,h_0),
\end{equation}
where $I_0$ and $h_0$ are the initial values of the obliquity and the longitude
of the spin axis and $I(t,I_0,h_0)$ is the obliquity at time $t$ for these
initial values, in the moonless system and in the system with exo-Moon.
If upon addition of the exo-Moon the range $\Delta$ decreases, then we call
the impact of the exo-Moon {\it stabilising}, while if the range
increases we call the impact {\it destabilising}.

We study rotation of the exo-Earth under the torque due to other bodies
following the approach of \cite{kp18,pk20} (see also \cite{mk81})
by expanding the Hamiltonian describing rotation of the exo-Earth in a power
series in the small parameter and applying time-averaging related to
the order-one frequencies $\bom$. Averaging over one or several fast variables
\cite{be66,be72,bks03,c15,lh17,p69,w75,slb19} is often
applied to study rotation of selestial bodies.
It may possibly reduce investigation of temporal evolution of the rotation
axis of a planet or a satellite into an integrable problem that has an
analytical solution, which was the case in \cite{kp18,pk20} where evolution
of obliquity of an exo-Earth in a system comprised of a star and planets was
studied.

The paper has the following structure:\\
In section~\ref{sec2} we recall the Hamilton equations for rotation of a rigid
body and averaging. The averaged equations involve six coefficients, which
are constants in the moonless system and become time-periodic as the
exo-Moon is added. The coefficients are computed given the masses and orbits
of the celestial bodies. In section~\ref{sec5} we calculate analytically
the range of obliquity in a system comprised of exo-Sun and
exo-planets, using the fact that the mass of exo-Sun is much larger
than the masses of the planets.
In section~\ref{sec6} we study the impact of the exo-Moon using the
results of section \ref{sec5}. We start by presenting
examples of systems where the impact of the exo-Moon is stabilising or
destabilising, which is proven analytically.
In what follows we consider
a simple system, comprised of a star and two planets, exo-Sun, exo-Earth and
exo-Jupiter, where
the orbits of the planets are given Keplerian ellipses (see Fig.\ref{fig1}).
We investigate numerically how the addition of an exo-Moon modifies the range of
nutation angle depending on the eccentricities, semi-major axes
and inclinations of the exo-Jupiter's and exo-Moon's orbits.
Finally, we briefly summarise
our results and indicate possible directions for the further studies.
In the appendix for completness of the presentation we calculate analytically
the range of obliquity in a system comprised of exo-Sun, exo-Earth and
exo-Moon, using the same approach as in section \ref{sec5}.

\section{Equations of motion}\label{sec2}

In this section we recall Hamiltonian equations for rotation of a rigid body
and apply averaging to derive the equations
that govern the behaviour of obliquity of the exo-Earth on large time scales.
The presentation follows \cite{pk20} where the evolution
of obliquity of a moonless exo-Earth was studied. Hence we skip some details
that can be found {\it ibid}.

\subsection{Hamiltonian equations}\label{sec21}

Denote by $OXYZ$ a non-moving inertial reference frame,
by $M\xi\eta\zeta$ the coordinate system whose
origin is at the center of mass of the exo-Earth and axes are parallel to those
of the $OXYZ$, and by $Mxyz$ the coordinate system with the same origin and
coordinate axes coinciding with the exo-Earth's principal axes. We assume that
$Mz$ is the axis associated with the maximum moment of inertia.

To investigate rotation of the exo-Earth we employ the Andoyer
variables \cite{an23}, for which following \cite{ki77}
we use the notation $(G,H,L,g,h,l)$, where\\
$G$ is the magnitude of the exo-Earth angular momentum vector $\bL$,\\
$H$ is the $Z$-component of $\bL$,\\
$L$ is the $z$-component of $\bL$,\\
$g$ is the angle between intersections of the plane $\Sigma$ with
the planes $M\xi\eta$ and $Mxy$,\\
$h$ is the angle between the axis $M\xi$ and the intersection of the
planes $\Sigma$ and $M\xi\eta$,\\
$l$ is the angle between the axis $Mx$ and the intersection of the
planes $\Sigma$ and $Mxy$\\
and $\Sigma$ the equatorial plane orthogonal to $\bL$.
The respective Hamilton equations for the rotating of axially symmetric
rigid exo-Earth then are
\begin{equation}\label{eqmo}
{\rd\over\rd t}(g,h,l)={\partial{\cal H}\over\partial(G,H,L)},\qquad
{\rd\over\rd t}(G,H,L)=-{\partial{\cal H}\over\partial(g,h,l)}
\end{equation}
where the Hamiltonian is
\begin{equation}\label{hamil}
{\cal H}=\frac{G^2-L^2}{2A}+\frac{L^2}{2C}+\sum_{n=1}^N{V_n},
\end{equation}
$A=B<C$ are the principal moments of inertia of the exo-Earth and $V_n$ is
the potential energy of the gravitational interaction
with the $n$-th celestial body, $N$ being the number of celestial bodies in the system,
other then exo-Earth.
Assuming that radius of a planet is small compare to the distance between
celestial bodies, only the leading-order part of the potential energy is
preserved, namely
\begin{equation}\label{grpt}
V_n=\frac{3\mu_n}{2R_n^3}(C-A)\beta_n^2,\quad\mu_n=fm_n,
\end{equation}
where $f$ is the universal gravitation constant, $m_n$ is the mass of the body,
$R_n$ is its geocentric distance, and $\beta_n$ is the cosine of the angle
between the directional vector $\bR_n=(R_{nX},R_{nY},R_{nZ})$ from the geocenter
to the $n$-th body and the $Mz$-axis.
For an axially symmetric body the r.h.s. of (\ref{hamil}) if independent
of $l$, therefore the $z$-component of the angular momentum vector does not
change in time.

\subsection{Averaging}\label{sec22}

The planets of the Solar system are almost spherical, hence it is natural
to assume that for the exo-Earth this also holds true. For a small
$\varepsilon=(C-A)/C\ll 1$, we rewrite the Hamiltonian (\ref{hamil}),
(\ref{grpt}) as
\begin{equation}\label{hamde}
\cH=\frac{G^2}{2J_0}+\varepsilon\cH_1+\ro(\varepsilon),
\end{equation}
where, by (\ref{grpt}),
\begin{equation}\label{ham1}
\cH_1=-\frac{1}{2J_0^2}\,\left[L^2C_1+(G-L^2) A_1\right]
+\frac{3}{2}\,\sum_{n=1}^N\frac{\mu_n}{R_n^3}(C_1-A_1)\beta_n^2,
\end{equation}
$J_0=(2A+C)/3$ is the mean moment of inertia of the exo-Earth,
\begin{equation}\label{a1c1}
A=J_0+\varepsilon A_1\hbox{ and }C=J_0+\varepsilon C_1.
\end{equation}

Let $\bom=(\omega,\omega_1,...,\omega_K)$ be the $K+1$ prescribed
order-one frequencies of motion of the considered $N+1$ celestial bodies,
i.e., any coordinate $Q(t)$
(where $Q$ stands for $X,Y,Z,X_n,Y_n$ or $Z_n$) can be expressed as
\begin{equation}\label{qsum}
Q(t)=\sum_{\bs=(s,s_1,...,s_K),\ 0<|\bs|<\infty}
q_{\bs}\re^{\ri(\bs\cdot\bom)t}.
\end{equation}
In the case of several fast frequencies, one can either employ the so-called
general averaging \cite{kr15,sa85,v62}, or following \cite{kz93,mk81}
introduce the fast variables
$$\theta=\omega t,\quad\theta_k=\omega_k t,\quad 1\le k\le K,$$
and define an average of a function $F$ as
\begin{equation}\label{avF}
\overline F=\frac{1}{(2\pi)^{K+2}}\int_0^{2\pi}...\int_0^{2\pi}
F\rd g\rd\theta\rd\theta_1\ldots\rd\theta_K.
\end{equation}

In the absence of resonances between $\omega$, $\omega_k$ and $G/J_0$,
\begin{equation}\label{avham}
\cH\approx\frac{G^2}{2J_0}+\varepsilon\overline\cH_1
\end{equation}
for a small $\varepsilon$, where by (\ref{grpt})-(\ref{a1c1}) the mean
Hamiltonian is
\begin{equation}\label{eqoH}
\overline\cH_1=\cF(G,L,l)\cG(G,H,h).
\end{equation}
Here
\begin{equation}\label{eqFG}
\renewcommand{\arraystretch}{1.5}
\begin{array}{l}
\cF(G,L,l)=-(C_1-A_1)(\displaystyle\frac{2}{3}-\sin^2 J),\\
\cG(G,H,h)=\displaystyle\frac{G^2}{2J_0^2}+
\displaystyle\frac{9}{4}\left[\left(\displaystyle\frac{1}{3}-
\cos^2h\sin^2 I\right)\sum_{n=1}^N\mu_n\left(D_{nY^2}-D_{nX^2}\right)\right.\\
-\left(\displaystyle\frac{2}{3}-\sin^2 I\right)
\sum_{n=1}^N\mu_n(D_{nZ^2}-D_{nX^2})+\sin(2h)\sin^2I\sum_{n=1}^N\mu_nD_{nXY}\\
\left.-\sin(2I)\left(\sin h\sum_{n=1}^N\mu_nD_{nXZ}-
\cos h\sum_{n=1}^N\mu_nD_{nYZ}\right)\right]\\
\cos I=\frac{H}{G},\quad\cos J=\frac{L}{G}
\end{array}\end{equation}
and
\begin{equation}\label{for0}
D_{n\rho^2}=\frac{1}{(2\pi)^{K+1}}\int_0^{2\pi}...\int_0^{2\pi}
\frac{R_{n\rho}^2}{R_n^5}d\theta\rd\theta_1...\rd\theta_K,\
D_{n\rho\nu}=\frac{1}{(2\pi)^{K+1}}\int_0^{2\pi}...\int_0^{2\pi}
\frac{R_{n\rho}R_{n\nu}}{R_n^5}d\theta\rd\theta_1...\rd\theta_K,
\end{equation}
where $\rho$ and $\nu$ denote $X,Y$ or $Z$.

Below we only consider the case when the rotation axis coincides with
the symmetry axis of the body. In such a case
the angular momentum $\bL=(L_\xi,L_\eta,L_\zeta)$ takes the form
\begin{equation}\label{comL}
\begin{aligned}
L_\xi=&G\sin h\sin I\\
L_\eta=&-G\cos h\sin I\\
L_\zeta=&G\cos I.
\end{aligned}\end{equation}
By (\ref{eqmo}) and (\ref{avham})-(\ref{eqFG}), the evolution
of the angles $h$ and $I$ satisfies the ODEs
\begin{equation}\label{gamav}
{\rd h\over\rd t}={3\varepsilon\over2}(C_1-A_1){1\over G\sin I}
{\partial{\tilde\cG}\over\partial I},\qquad
{\rd I\over\rd t}=-{3\varepsilon\over2}(C_1-A_1){1\over G\sin I}
{\partial{\tilde\cG}\over\partial h},
\end{equation}
where
\begin{equation}\label{aveG2}
\tilde\cG(G,I,h)=(-D_1\sin^2h-D_2\cos^2h+D_3+D_4\sin(2h))\sin^2I
-\sin(2I)(D_5\sin h-D_6\cos h).
\end{equation}
Here, $D_j$ are coefficients,
\begin{equation}\label{coed}
D_j=\sum_{n=1}^N D_j^{(n)},\quad 1\le j\le 6,
\end{equation}
\begin{equation}\label{coed6}
\begin{array}{l}
D_1^{(n)}=m_nD_{nX^2},\quad D_2^{(n)}=m_nD_{nY^2},\quad D_3^{(n)}=m_nD_{nZ^2},\\
D_4^{(n)}=m_nD_{nXY},\quad D_5^{(n)}=m_nD_{nXZ},\quad D_6^{(n)}=m_nD_{nYZ};
\end{array}\end{equation}
and $D_{n\rho\nu}$ are given by (\ref{for0}).
In sums (\ref{coed}) the terms $D_j^{(n)}$, $1\le j\le 6$, originate from
the gravitation interaction of the exo-Earth with the $n$-th celestial body.
We label the bodies as follows: the first one is exo-Sun, the second is
exo-Moon and the numbers from three to $n$ are attributed to exo-planets other
than exo-Earth.

\subsection{Calculation of coefficients $D_j$ related to the
exo-Moon and exo-Sun}\label{sec23}

The coefficients $D_j$ related to the planets, in general, should be
found numerically. Recall that we assume that the orbit of exo-Moon is a
Keplerian ellipse with a constant inclination to the ecliptic undergoing two
types of precessional
motion with respective frequencies $\sigma_n$ and $\sigma_a$.
In such a case the coefficients $D_j^{(2)}$ related to the exo-Moon can
be found analytically and we evaluate them in this subsection.

If the orbit of the exo-Earth around exo-Sun is a Keplerian ellipse then
then the respective coefficients $D_j^{(1)}$ can be calculated as well.
The plane $OXY$ being the orbital plane of the exo-Earth, its elliptic orbit
satisfies the relations
\begin{equation}\label{trap}
\renewcommand{\arraystretch}{2.}
\begin{array}{l}
X=\displaystyle{\frac{a_\rE(1-e_\rE)^2\cos\nu_\rE}{1+e_\rE\cos\nu_\rE}},
\quad Y=\displaystyle{{\frac{a_\rE(1-e_\rE)^2\sin\nu_\rE}{1+e_\rE\cos\nu_\rE}}},\quad
Z=0\\
\end{array}\end{equation}
where $a_\rE$, $e_\rE$ and $\nu_\rE$ are the semi-major axis, essentricity and
the true anomaly of the
exo-Earth. Following the averaging procedure discussed
in \cite{pk20}, we introduce the fast variable $\theta_\rE=\omega_\rE t$,
where $\theta_\rE$ is the mean anomaly of the exo-Earth's orbits, that is
related to the true anomaly as follows:
\begin{equation*}
\frac{d\theta_\rE}{d\nu_\rE}=\frac{(1-e_\rE^2)^{3/2}}{(1+e_\rE\cos\nu_\rE)^2}.
\end{equation*}
Since $X_1(t)=Y_1(t)=Z_1(t)=0$, the coefficients are:
\begin{equation}\label{coE}
D_1^{(1)}=D_2^{(1)}={\mu_\rS(1-e_\rE^2)^{3/2}\over 2p_\rE^3},\
D_3^{(1)}=D_4^{(1)}=D_5^{(1)}=D_6^{(1)}=0.
\end{equation}

The orbit of the exo-Moon is an ellipse with exo-Earth being one of the focuses.
The inclination $i$ of the lunar orbit to
the ecliptic plane does not change in time. Longtitude of the ascending node
and the argument of periapsis evolve as
$$\Omega=\Omega_0+\sigma_a t,\quad \omega=\omega_0+\sigma_nt.$$
To calculate the coefficients $D_j^{(2)}$ recall that (see, e.g. \cite{balk})
\begin{equation}\label{eq30}
\renewcommand{\arraystretch}{1.5}
\begin{array}{ll}
R_{2X}=X-X_2=&[\cos\Omega\cos\omega-\sin\Omega\cos i\sin\omega]\xi'-\\
&[\cos\Omega\sin\omega+\sin\Omega\cos i\cos\omega]\eta'+\sin\Omega\sin i\zeta'\\
R_{2Y}=Y-Y_2=&[\sin\Omega\cos\omega+\cos\Omega\cos i\sin\omega]\xi'+\\
&[-\sin\Omega\sin\omega+\cos\Omega\cos i\cos\omega]\eta'-\cos\Omega\sin i\zeta'\\
R_{2Z}=Z-Z_2=&\sin i\sin\omega\xi'+\sin i\cos\omega\eta'+\cos i\zeta',
\end{array}\end{equation}
where $(\xi',\eta',\zeta')$ are the coordinates of the exo-Moon in the
coordinate system $M\xi',\eta',\zeta'$ related to the Moon's orbit:
the origin is in the center of mass of the exo-Earth, the positive $M\xi'$ axis
points to the perigee of the exo-Moon's orbit, the axis $M\eta'$ belongs
to the orbit and is obtained rotating the $M\xi'$ axis by $\pi/2$ in the
direction of the Moon's motion, the axis $M\zeta'$ is orthogonal to the
orbit and its direction is chosen to obtain a right-handed coordinate system.
The coordinates of a point in the orbit satisfy
$$\xi'={a_2(1-e_2^2)\over 1+e_2\cos\nu_2}\cos\nu_2,\quad
\eta'={a_2(1-e_2^2)\over 1+e_2\cos\nu_2}\sin\nu_2,\quad\zeta'=0,$$
$a_2$, $e_2$ and $\nu_2$ beinge the semi-major axis, essentricity and
true anomaly of the exo-Moon orbiting the exo-Earth. Therefore
\begin{equation}\label{eq31}
\renewcommand{\arraystretch}{1.5}
\begin{array}{l}
R_{2X}=\displaystyle{a_2(1-e_2^2)[\cos\Omega\cos(\omega+\nu_2)-
\sin\Omega\cos i\sin(\omega+\nu_2)]\over 1+e_2\cos\nu_2}\\
R_{2Y}=\displaystyle{a_2(1-e_2^2)[\sin\Omega\cos(\omega+\nu_2)+
\cos\Omega\cos i\sin(\omega+\nu_2)]\over 1+e_2\cos\nu_2}\\
R_{2Z}=\displaystyle{a_2(1-e_2^2)\sin i\sin(\omega+\nu_2)\over 1+e_2\cos\nu_2}.
\end{array}\end{equation}
Substituting (\ref{eq31}) into (\ref{for0}) and (\ref{coed6}) we obtain that
\begin{equation}\label{eq32}
\begin{array}{l}
\renewcommand{\arraystretch}{1.5}
D_1^{(2)}=\Xi(\cos^2\Omega\sin^2i+\cos^2i),\quad
D_2^{(2)}=\Xi(-\cos^2\Omega\sin^2i+1),\quad
D_3^{(2)}=\Xi\sin^2i\\
D_4^{(2)}=\Xi\sin\Omega\cos\Omega\sin^2i,\quad
D_5^{(2)}=-\Xi\sin\Omega\sin i\cos i,\quad
D_6^{(2)}=\Xi\cos\Omega\sin i\cos i,
\end{array}\end{equation}
where $\Omega=\Omega_0+\sigma_at$ and
\begin{equation}\label{eq32a}
\Xi={fm_2\over 2a_2^3(1-e_2^2)^{3/2}}.
\end{equation}

\section{Planetary system, comprised of exo-Sun, exo-Earth and exo-planets.}
\label{sec5}

Evolution of the obliquity in the system considered in this section was studied
in general setup in \cite{pk20}. Here we derive an approximation
for the range of obliquity using the fact that the mass of the star is much
larger than the mass of any planet.

In a system comprised of the exo-Sun and exo-planets only
the evolution of the angles $h$ and $I$ satisfies the ODEs
(\ref{gamav}),(\ref{aveG2}) with $D_j$, $j=1,...,6$, being time-independent
constants.
The equation (\ref{aveG2}) is invariant under the symmetry
$(I,h)\to(\pi-I,h+\pi)$.

The right hand side of (\ref{aveG2}) can be represented as
\begin{equation}\label{aveG200}
\tilde\cG(I,h)=\sin^2I(-D+\alpha\cos2h')+\beta\sin2I\cos(h'+\gamma),
\end{equation}
where
\begin{equation}\label{aveG222}
D={D_1+D_2-2D_3\over 2},\
\alpha=\biggl({(D_1-D_2)^2\over4}+D_4^2\biggr)^{1/2},\
\beta=(D_5^2+D_6^2)^{1/2},
\end{equation}
$h'=h+\arctan(2D_4/(D_1-D_2))/2$ and
$\gamma=\arctan(D_6/D_5)-\arctan(2D_4/(D_1-D_2))/2$.

Steady states of system (\ref{gamav}),(\ref{aveG200}) can be found from
the equations
\begin{equation}\label{aveGGG}
\renewcommand{\arraystretch}{1.5}
\begin{array}{l}
\sin2I(-D+\alpha\cos2h')+\beta\cos2I\cos(h'+\gamma)=0,\\
-\sin^2I\alpha\sin2h'-\beta\sin2I\sin(h'+\gamma)=0.
\end{array}
\end{equation}

Since the mass of the exo-Sun is much larger than the masses of planets,
due to (\ref{coed}), (\ref{coed6}) and (\ref{aveG222}), we have
$D\gg\max(|\alpha|,|\beta|)$. Therefore the first equation in (\ref{aveGGG})
implies $\sin2I\approx0$. Hence, the steady states are
$$I\approx 0;\quad I\approx \pi;\quad
I\approx \pi/2,\ h'\approx0,\pi/2,\pi,3\pi/2,$$
where those $(I\approx\pi/2,h'\approx0)$ and $(I\approx\pi/2,h\approx\pi)$
are saddles and the other ones are centers. The saddles are connected by
heteroclinic trajectories that divide
the celestial sphere into four regions, each comprised of a center and
closed trajectories around it, see Fig.~\ref{fig3}a. We call {\it polar} the regions around
steady states $I\approx0$ and $\pi$ and {\it equatorial} the ones around steady
states with $I\approx\pi/2$.
Since $\rd I/\rd h'=\rd I/\rd t(\rd h'/\rd t)^{-1}$, the extrema of $I(h')$ take
place at
\begin{equation}\label{exxx}
-\sin^2I\alpha\sin2h'-\beta\sin2I\sin(h'+\gamma)=0.
\end{equation}

Below we evaluate $\Delta(I_0,h_0)$ defined by (\ref{Delta}), where $h_0=\pi/2$
and consider $0\le I_0\le\pi/2$. Since $D$ is large compare with $\alpha$
and $\beta$, the obliquity $I(h')$ is close to $I_0$ and we can write
$I(h')=I_0+I_1(h')$. Therefore,
$$\sin^2I\approx\sin^2I_0+I_1\sin2I_0+I_1^2\cos2I_0,\quad
\sin2I\approx\sin2I_0+2I_1\cos2I_0.$$
Since $\tilde\cG(I,h')$ is a constant on trajectories, substituting the above
expressions into (\ref{aveG200}) we obtain a quadratic equation on $I_1(h')$
\begin{equation}\label{quaI1}
\renewcommand{\arraystretch}{1.5}
\begin{array}{l}
I_1^2(-D\cos2I_0)+I_1(-D\sin2I_0+2\beta\cos2I_0\cos(h'+\gamma))+\\
\alpha\sin^2I_0(\cos2h'-1)+\beta\sin2I_0(\cos(h'+\gamma)+\sin\gamma)=0.
\end{array}
\end{equation}

The range $\Delta$ is a continuous function of $I_0$ and $h_0$ inside a region
and is discontinuous at a boundary. For a
heteroclinic trajectory through $(\pi/2,0)$ we have
$$\tilde\cG(I,h)=(-D+\alpha),$$
hence trajectories with the initial conditions $(I_0,\pi/2)$ such that
\begin{equation}\label{equat}
\sin^2I_0(-D-\alpha)+\beta\sin2I_0\sin\gamma<-D+\alpha
\end{equation}
belong to the equatorial regions, while the other ones to the polar regions.
Since $D\gg\max(\alpha,\beta)$ the inequality (\ref{equat}) may be simplified to
\begin{equation}\label{equatS}
|I_0-{\pi\over 2}|<\biggl({2\alpha\over D}\biggr)^{1/2},
\end{equation}
i.e. the trajectories through $(I_0,\pi/2)$ belong to an equatorial region
if $\pi/2-\delta_{het}<I_0<\pi/2+\delta_{het}$, where
$\delta_{het}=(2\alpha/D)^{1/2}$, and to a polar one otherwise.

To solve the equation (\ref{quaI1}) we regard three possibilities for $I_0$
if the initial condition $(I_0,h_0)$ belongs to the polar region:
\begin{itemize}
\item [(i)] $D^{1/2}|\sin I_0|<\beta^{1/2}$, $|\sin I_0|<|\cos I_0|$;
\item [(ii)] $D^{1/2}\sin^2I_0>\max(\alpha,\beta^{1/2}|\sin I_0|)$;
\item [(iii)] $D^{1/2}\sin^2I_0<\alpha^{1/2}$, $|\sin I_0|>|\cos I_0|$,
$|I_0-{\pi\over 2}|>(2\alpha)^{1/2}D^{-1/2}$.
\end{itemize}
and separately consider the equatorial region
\begin{itemize}
\item [(iv)] $|I_0-{\pi\over 2}|<(2\alpha)^{1/2}D^{-1/2}$.
\end{itemize}

Since $D\gg\beta$, in case (i) we have that $I_0$ is close to 0 or $\pi$.
Therefore, from (\ref{exxx}) the extrema of $I(h')$ take place at
$\sin(h'+\gamma)\approx 0$. Substituting $h'+\gamma=0$ and $\pi$ into (\ref{quaI1}),
solving the quadratic  equation and subtracting the root at $h'+\gamma=0$
from the one at $h'+\gamma=\pi$ we find that
\begin{equation}\label{casei}
\Delta\approx {2\beta\over D}.
\end{equation}

In case (ii) in the equation (\ref{quaI1}) the quadratic term can be neglected
and the remaining linear equation can be easily solved for any value of $h'$.
We can not derive from (\ref{exxx}) the particular value of $h'$ where the
maxima and minima take place, hence we can give upper and lower bound for
$\Delta$ (they differ less than a factor 2):
\begin{equation}\label{caseii}
{\max(|\alpha|\sin I_0,2|\beta|\cos I_0)\over D\cos I_0}<\Delta\le
{|\alpha|\sin I_0+2|\beta|\cos I_0\over D\cos I_0}.
\end{equation}
Alternatively, we introduce the function
$$\chi(I_0,\alpha,\beta,\gamma)=
\max_{0\le h\le 2\pi}f(I_0,\alpha,\beta,\gamma,h)-
\min_{0\le h\le 2\pi}f(I_0,\alpha,\beta,\gamma,h)$$
where
$$f(I_0,\alpha,\beta,\gamma,h)=\alpha\sin^2I_0\cos2h+\beta\sin2I_0\cos(h+\gamma).$$
Then
\begin{equation}\label{caseiia}
\Delta(I_0,h_0)\approx{\chi(I_0,\alpha,\beta,\gamma)\over D\cos I_0}.
\end{equation}

As we noted $D\gg\alpha$, therefore in case (iii) we have that
$I_0\approx\pi/2$. Hence (see (\ref{exxx})\,) the minima of $I(h)$ take place
at $h=0$ and $\pi$ and the maxima at $h=\pi/2$ and $3\pi/2$. Solving
(\ref{quaI1}) we obtain that
\begin{equation}\label{caseiii}
\Delta\approx{D\sin2I_0-\beta\cos2I_0\cos\gamma+
((-D\sin2I_0+\beta\cos2I_0\cos\gamma)^2+8D\alpha)^{1/2}\over2D}.
\end{equation}

In case (iv) a trajectory twice intersects the meridian $h=\pi/2$,
at the intersection points $I$ takes the maximum and minimum values,
$I_{\max}$ and $I_{\min}$, for this particular trajectory (see (\ref{exxx})\,).
Moreover, (\ref{quaI1}) implies that $I_{\max}-\pi/2\approx\pi/2-I_{\min}$.
Hence,
\begin{equation}\label{caseiv}
\Delta\approx 2|\pi/2-I_0|.
\end{equation}

Overall, for $0\le I_0\le\pi/2$ we have
\begin{equation}\label{DeltPl}
\renewcommand{\arraystretch}{1.5}
\begin{array}{lll}
\hbox{ case } & I_0 & \Delta\\
\hline
(i) & D^{1/2}|\sin I_0|<\beta^{1/2} & \hbox{ equation (\ref{casei}) }\\
(ii) & D^{1/2}\sin^2I_0>\max(\alpha,\beta^{1/2}|\sin I_0|) &
\hbox{ inequality (\ref{caseii}) }\\
&&\hbox{ or equation (\ref{caseiia}) }\\
(iii) & D^{1/2}\sin^2I_0<\alpha^{1/2},\
|I_0-{\pi\over 2}|>(2\alpha)^{1/2}(D')^{-1/2} & \hbox{ equation (\ref{caseiii}) }\\
(iv) & |I_0-{\pi\over 2}|<(2\alpha)^{1/2}D^{-1/2} &
\hbox{ equation (\ref{caseiv}) }
\end{array}\end{equation}

\begin{figure}[h]
{\large
\vspace*{1mm}
\hspace*{-15mm}\includegraphics[width=8cm]{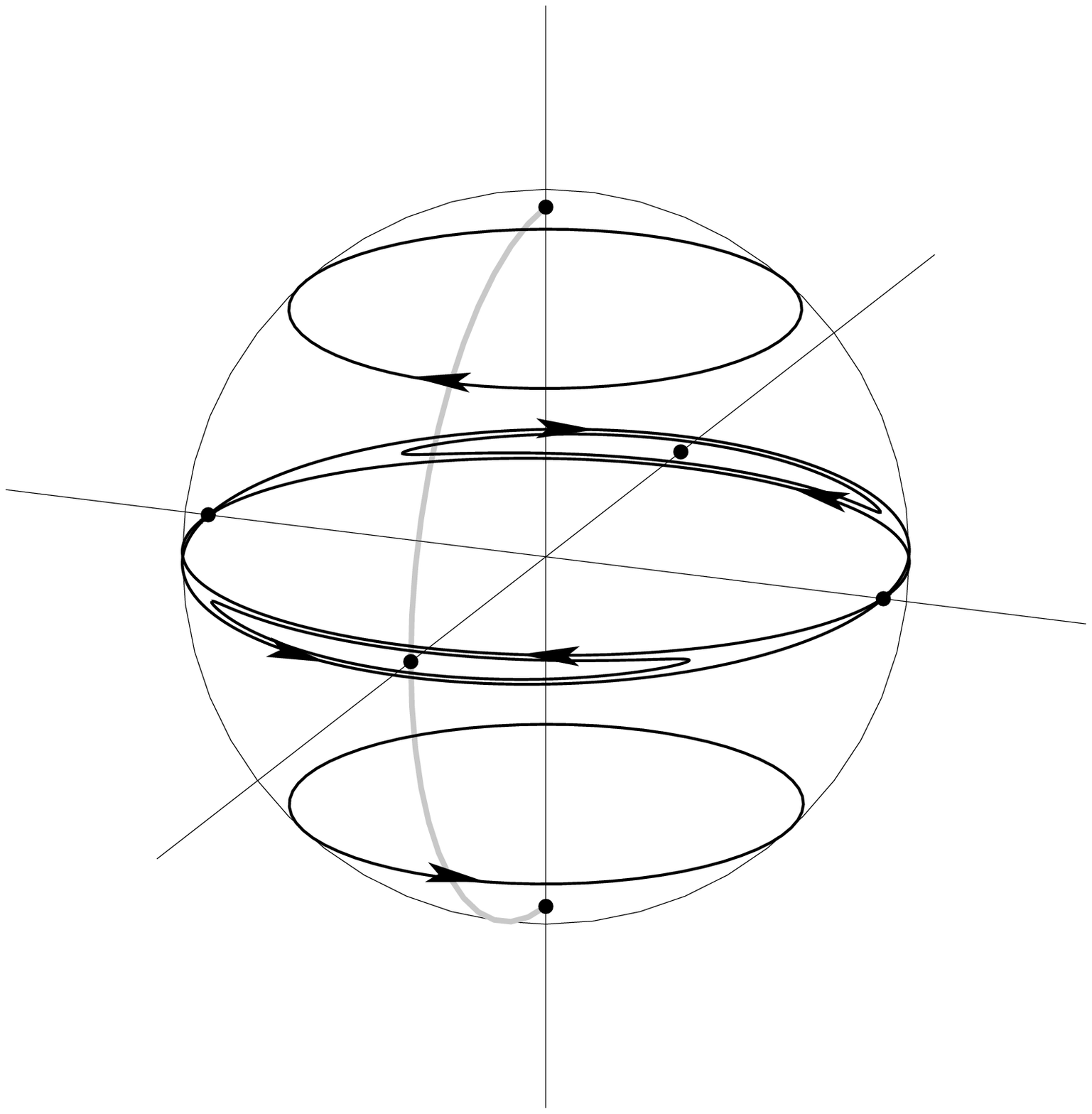}
\vspace*{4mm}

\vspace*{-49mm}
\hspace*{-6mm}$\xi$

\vspace*{13mm}
\hspace*{8mm}$\eta$

\vspace*{-56mm}
\hspace*{23mm}$\zeta$

\hspace*{75mm}\includegraphics[width=8.5cm]{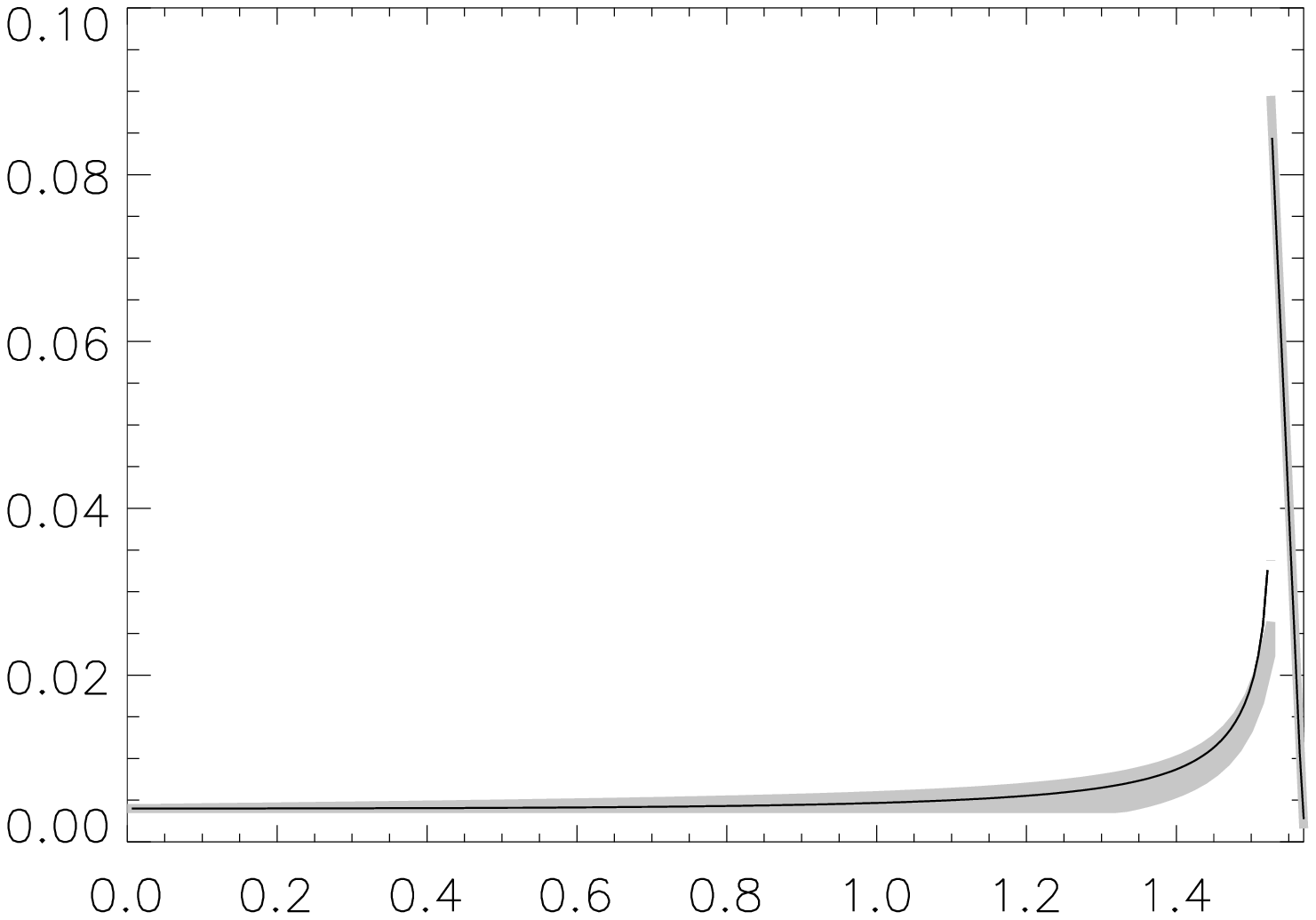}

\vspace*{-45mm}
\hspace*{73mm}{\large $\Delta$}

\vspace*{35mm}
\hspace*{131mm}{\large $I_0$}

\vspace*{5mm}
\hspace*{40mm}(a)\hspace*{82mm}(b)

}
\noindent
\caption{
Motion of $\bL$ (\ref{comL}) on the celestial sphere computed from
(\ref{gamav}), (\ref{aveG2}) (a) for $D=1$, $\alpha=0.001$, $\beta=0.002$
and $\gamma=\pi/8$
and $\Delta(I_0,\pi/2)$ as a function of $I_0$ calculated by integrating
the equations (\ref{gamav}),(\ref{aveG2}) (black line) and from the
approximations (\ref{DeltPl}) (gray area), (b).
The meridian $h=\pi/2$ is shown by gray line on the sphere.}
\label{fig2}\end{figure}

\begin{remark}\label{rem1}
We have calculated $\Delta(I_0,h_0)$ for $h_0=\pi/2$ only.
Unless $I_0$ is close to $\pi/2$, the range $\Delta$ if
independent from $h_0$, see approximations (\ref{casei}) and (\ref{caseiia}).
By contrast, near $I_0=\pi/2$ the range essentially depends on $h_0$, as
it can be see in fig.~\ref{fig2}a. In particular, it vanishes only
at meridians $h_0=\pi/2$ and $3\pi/2$, while the meridians $h_0=0$ and
$\pi$ do not cross the equatorial region. Investigation of the
dependence of $\Delta(I_0,h_0)$ on $h_0$ for $I_0$ near $\pi/2$, which can
carried out similarly, is left for future studies.
\end{remark}

\begin{figure}[h]
\vspace*{1mm}
\hspace*{-12mm}\includegraphics[width=8cm]{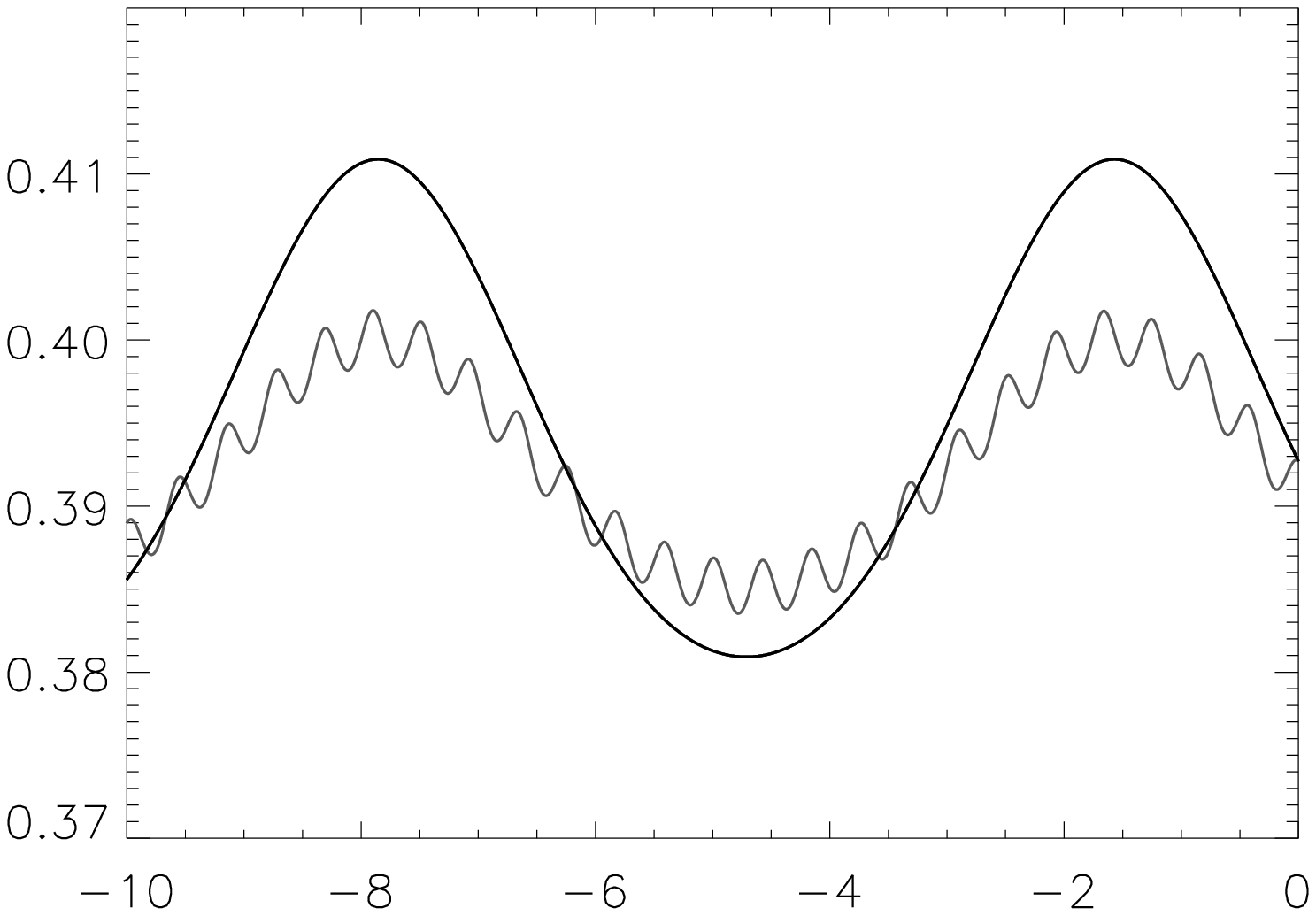}
\hspace*{10mm}\includegraphics[width=8cm]{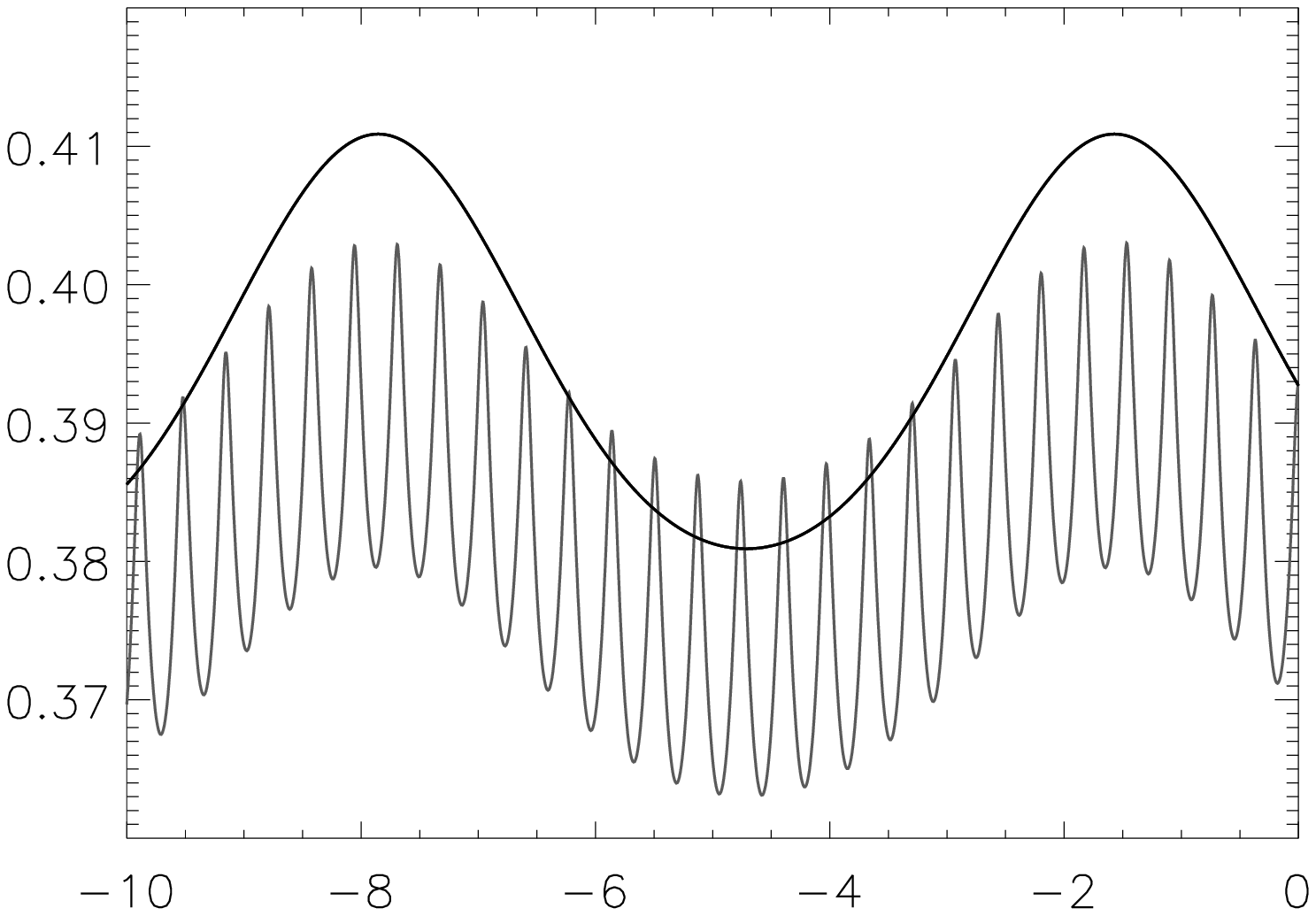}

\vspace*{-45mm}
\hspace*{-10mm}{\large $I$\hspace*{89mm}$I$}

\vspace*{37mm}
\hspace*{32mm}{\large $h$\hspace*{88mm}$h$}

\vspace*{1mm}
\hspace*{60mm}{\large (a)\hspace*{82mm}(b)}

\vspace*{3mm}
\noindent
\caption{The dependence of $I$ on $h$ for a moonless system (black line)
and for such system with added exo-Moon (gray line).
The parameters of the planetary system are: $m_\rS=1$, $a_\rE=1$, $e_\rE=0$,
$m\rJ=0.05$, $a_\rJ=1.5$, $e_\rJ=0.1$, $i=\pi/64$ (a) and $i=\pi/8$ (b).
The initial condition is $(I_0,h_0)=(\pi/8,0)$.}
\label{fig3}\end{figure}

\section{Planetary system, comprised of exo-Sun, exo-Earth, exo-Moon and
exo-planets.}
\label{sec6}

In this section we study how the range of obliquity changes as we add an
exo-Moon to the system considered in previous section.
As  it is shown in Fig.~\ref{fig3}, the addition of exo-Moon may result on decrease or increase of the
range.
In subsection \ref{sec61} we prove analytically that for certain systems the
impact of the exo-Moon is stabilising while in subsection \ref{sec62}
we prove that for some systems it is destabilising.
In subsection \ref{sec63} we study numerically the impact of
the exo-Moon in a particular system comprised of exo-Sun, exo-Earth and
an exo-planet as the orbital parameters of the exo-Moon and exo-planet are
varied.

\subsection{Stabilising moon.}
\label{sec61}

Let $\Delta^P(I_0,h)$ denotes the range of $I$ in the moonless system considered
in section \ref{sec5} and $\Delta^{P+M}(I_0,h)$ the range in the system with
added exo-Moon. Denote by $D_j^{P+M}$ the coefficients of equations
(\ref{gamav}),(\ref{aveG2}) in the system equipped with exo-Moon and by $D_j^P$
the coefficients in the moonless system. In agreement with (\ref{coed}) we have
that $D_j^{P+M}=D_j^{P}+D_j^{(2)}$, $j=1,...,6$. Suppose that the orbit of
the exo-Moon belongs to the ecliptic, i.e. $i=0$. Then the coefficients
$D_j^{(2)}$ (see (\ref{eq32})\,) satisfy
\begin{equation}\label{eq33}
D_1^{(2)}=D_2^{(2)}>0,\ D_3^{(2)}=D_4^{(2)}=D_5^{(2)}=D_6^{(2)}=0,
\end{equation}
where $D_1^{(2)}$ is time-independent.

Therefore results of section \ref{sec5} can be applied to the modified
system. Moreover, we have that $D^{P+M}>D^P$, $\alpha^{P+M}=\alpha^P$
and $\beta^{P+M}=\beta^P$ (see (\ref{aveG222})\,), where the upper indices refer
to the original system or to the system with added exo-Moon. If $I_0$
satisfies (i) or (ii) in (\ref{DeltPl}) then the respective expressions
for $\Delta$ involve $D$ in the denumeralor only, which implies that
$\Delta^{P+M}(I_0,h_0)<\Delta^P(I_0,h_0)$ unless $I_0$ is close to $\pi/2$.
The stabilising influence of exo-Moon is illustrated by Fig.~\ref{fig4}a.
Near $I_0=\pi/2$ the range essentially depends on $h_0$ (see remark \ref{rem1}),
which indicates that the type of impact should also depend on $h_0$.

\begin{remark}\label{rem2}
If the orbits of exo-Earth and exo-Moon were circular then $\sigma_a$ is
\cite{smart}
$$\sigma_a=-{3\over 4}{\omega_\rE^2\over\omega_\rM},$$
i.e., it is also order one, as $\omega_\rE$ and $\omega_\rM$.
The property is likely to hold true for orbits that are not very different
from circular. Suppose that in a system with $i\ne0$ the averaging over the
fast variable $\sigma_at$ is performed as well. Then coefficients (\ref{eq32})
related to the exo-Moon become time-independent and safisfy
\begin{equation}\label{eq34}
D_1^{(2)}=D_2^{(2)}>0,\ D_3^{(2)}>0,\ D_4^{(2)}=D_5^{(2)}=D_6^{(2)}=0.
\end{equation}
By the same arguments as applied above (\ref{eq34}) implies that
$\Delta^{P+M}(I_0,h_0)<\Delta^P(I_0,h_0)$ for $I_0$ which is not close to
$\pi/2$. Therefore, the impact of the exo-Moon is stabilising for amost all initial
conditions.
\end{remark}

\subsection{Destabilising moon.}
\label{sec62}

Consider a moonless system where the orbit of all planets including exo-Earth
are circular and belong to the equatorial plane. In such a system the
coefficients $D_i\equiv D_i^{P}$ satisfy
\begin{equation}\label{eqdes}
D_1^{P}=D_2^{P}>0,\ D_3^{P}=D_4^{P}=D_5^{P}=D_6^{P}=0.
\end{equation}
The rotation of the planet reduces to a regular precession about the axis
orthogonal to its orbital plane. (See section 5 in \cite{pk20}.)
We have therefore $\Delta^P(I_0,h)=0$ for any initial condition.
As exo-Moon is added to the system, the respectively modified coefficients
$D_j^{P+M}$ are the same as in appendix \ref{sec4} with $D_j^{(1)}$ replaced
by $D_j^{P}$. Hence, the results of the appendix are applicable to
the full system and the range of nutation angle, except for some special
initial conditions, is positive. Since $\Delta^{P+M}>\Delta^{P}=0$ the impact
of the exo-Moon is destabilising. This case is illustrated by Fig.~\ref{fig5}a.

\subsection{Numerical study of the impact.}
\label{sec63}

In this subsection we investigate how an added exo-Moon affects the range of
nutation angle in a simple system composed of the exo-Sun, exo-Earth and
a planet that we call exo-Jupiter (see fig.~\ref{fig1}). We assume
that the exo-Sun is the origin of the $OXYZ$ coordinate system and the orbit of
exo-Earth belongs to the $OXY$ plane. The orbits of the exo-Earth
and exo-Jupiter are Keplerian ellipses, whose semi-major axes are $a_\rE$ and
$a_\rJ$, and eccentricities $e_\rE$ and $e_\rJ$, respectively, and the angle
between the orbital planes is $\gamma_\rJ$. The orbital planes intersect
along the axis $OY$, and the major axes of both ellipses are orthogonal
to $OY$.

Upon the canonical change of variables employed in appendix \ref{sec4}
the equations of motion (\ref{gamav}),(\ref{aveG2}) take the form
\begin{equation}\label{gamavn}
{\rd h\over\rd t}={\rho\over\sin I}{\partial{\tilde\cG'}\over\partial I},\qquad
{\rd I\over\rd t}=-{\rho\over\sin I}{\partial{\tilde\cG'}\over\partial h},
\end{equation}
\begin{equation}\label{aveG2n}
\renewcommand{\arraystretch}{1.5}
\begin{array}{l}{\displaystyle
\tilde\cG'(G,I,h)=(-D_1^{P}\sin^2h-D_2^{P}\cos^2h+D_3^{P}+
D_4^{P}\sin(2h)+\Xi({1\over2}\sin^2i-\cos^2i)+{\Xi\over2}\sin^2i)\sin^2I-}\\
{\displaystyle(D_5^{P}\sin h-D_6^{P}\cos h+{\Xi\over2}\sin2i\cos h)\sin2I+
\sigma\cos I},
\end{array}
\end{equation}
where
$$\rho={3\varepsilon(C_1-A_1)\over2G}\hbox{ and }\sigma={\sigma_a\over\rho}.$$
Equation (\ref{aveG2n}) indicates that the contribution of the exo-Moon into
the motion of rotation axis is determined by $i$, $\Xi$ and $\sigma$.

\begin{figure}[p]
\vspace*{-4mm}
\hspace*{-12mm}\includegraphics[width=8cm]{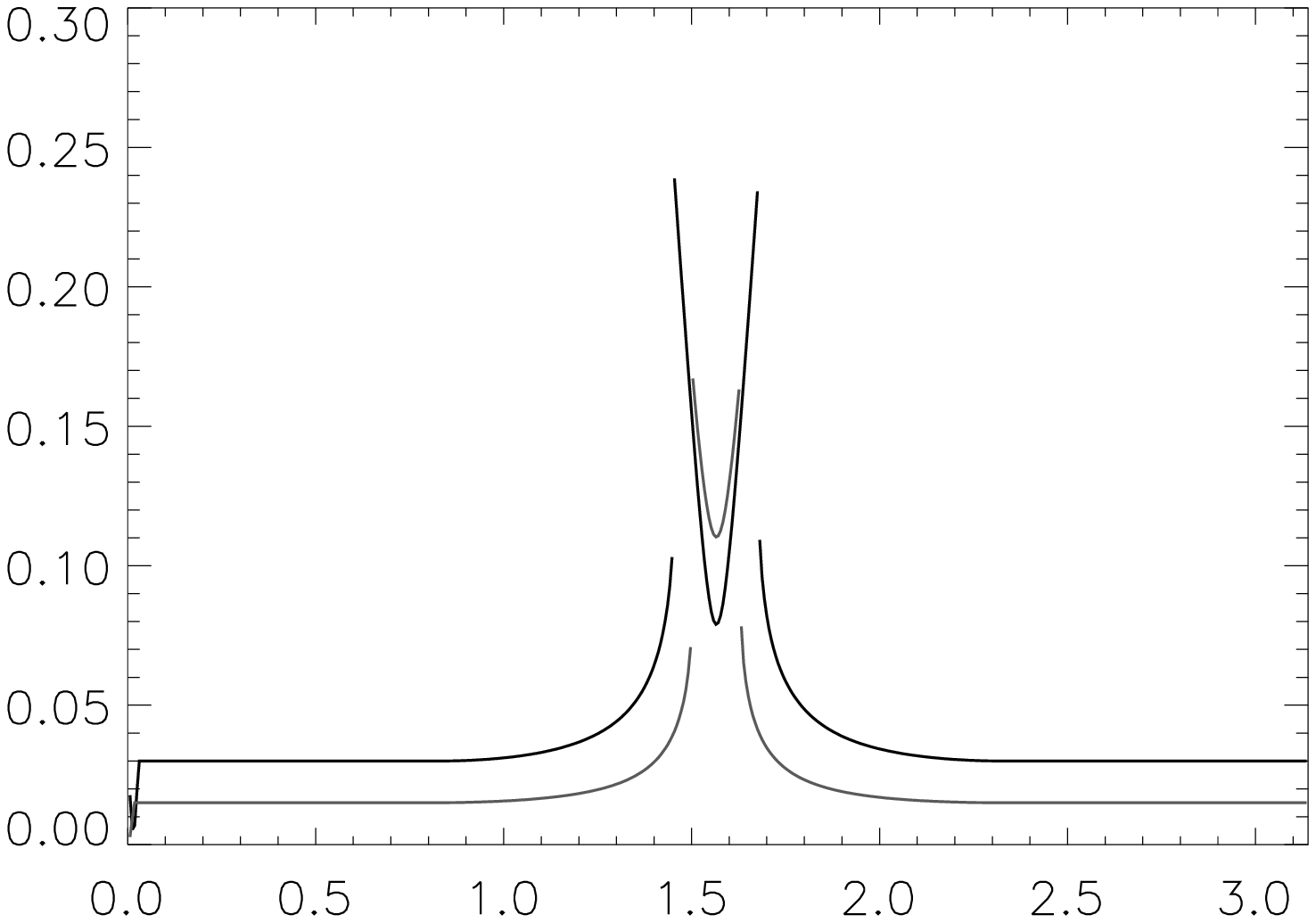}
\hspace*{10mm}\includegraphics[width=8cm]{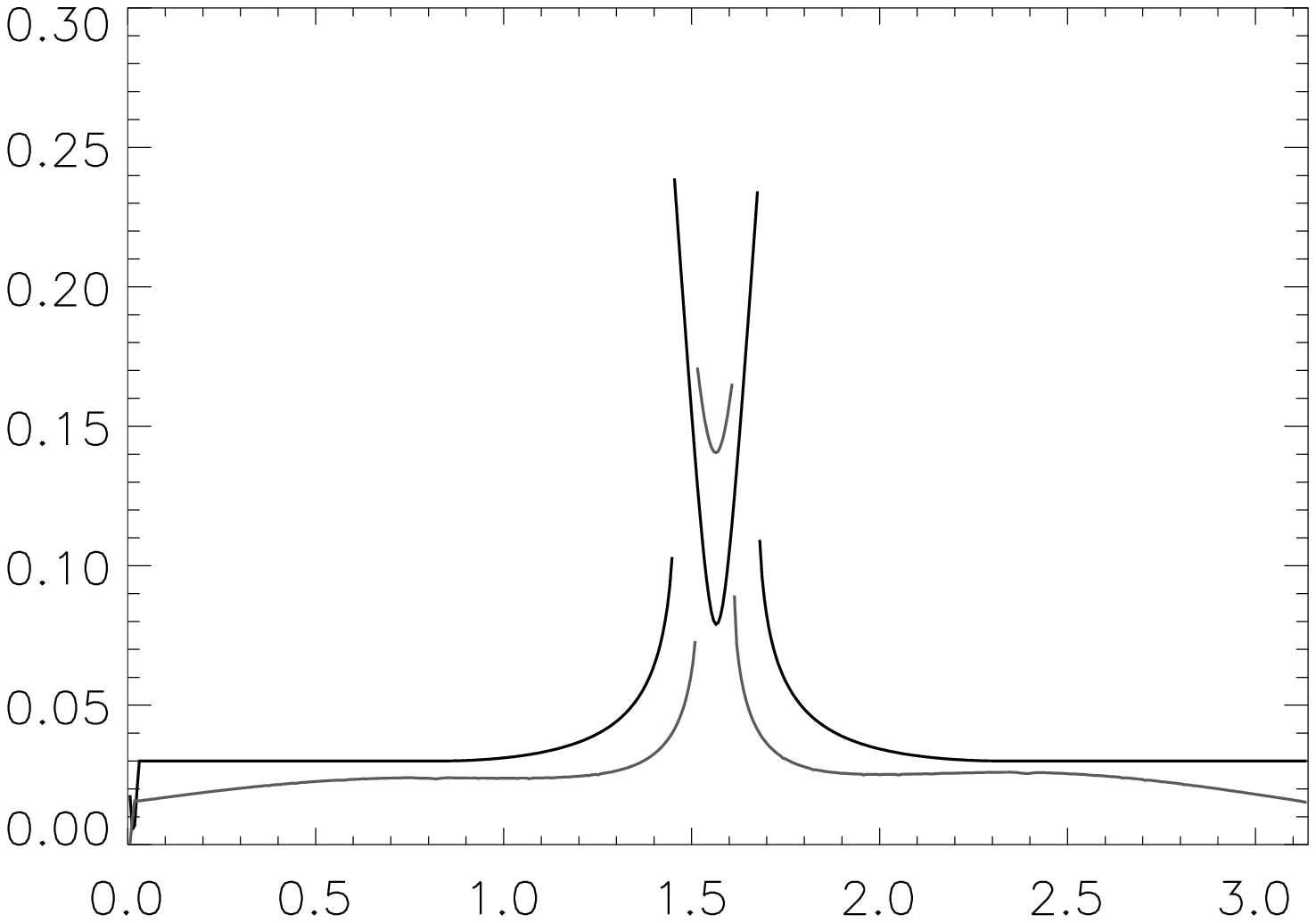}

\vspace*{-45mm}
\hspace*{-12mm}{\large $\Delta$\hspace*{87mm}$\Delta$}

\vspace*{37mm}
\hspace*{32mm}{\large $I_0$\hspace*{88mm}$I_0$}

\vspace*{1mm}
\hspace*{60mm}{\large (a)\hspace*{82mm}(b)}

\vspace*{4mm}
\hspace*{-12mm}\includegraphics[width=8cm]{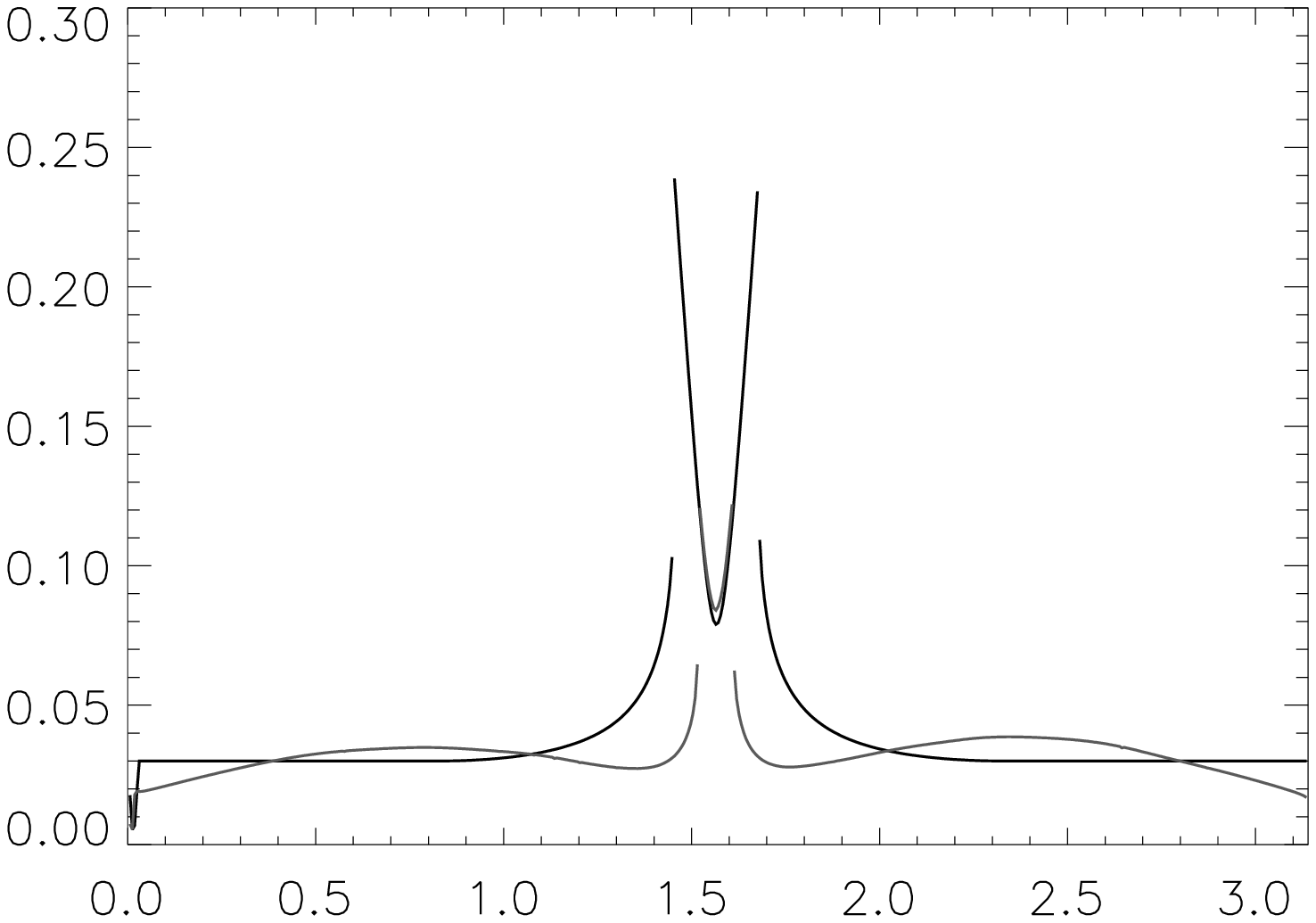}
\hspace*{10mm}\includegraphics[width=8cm]{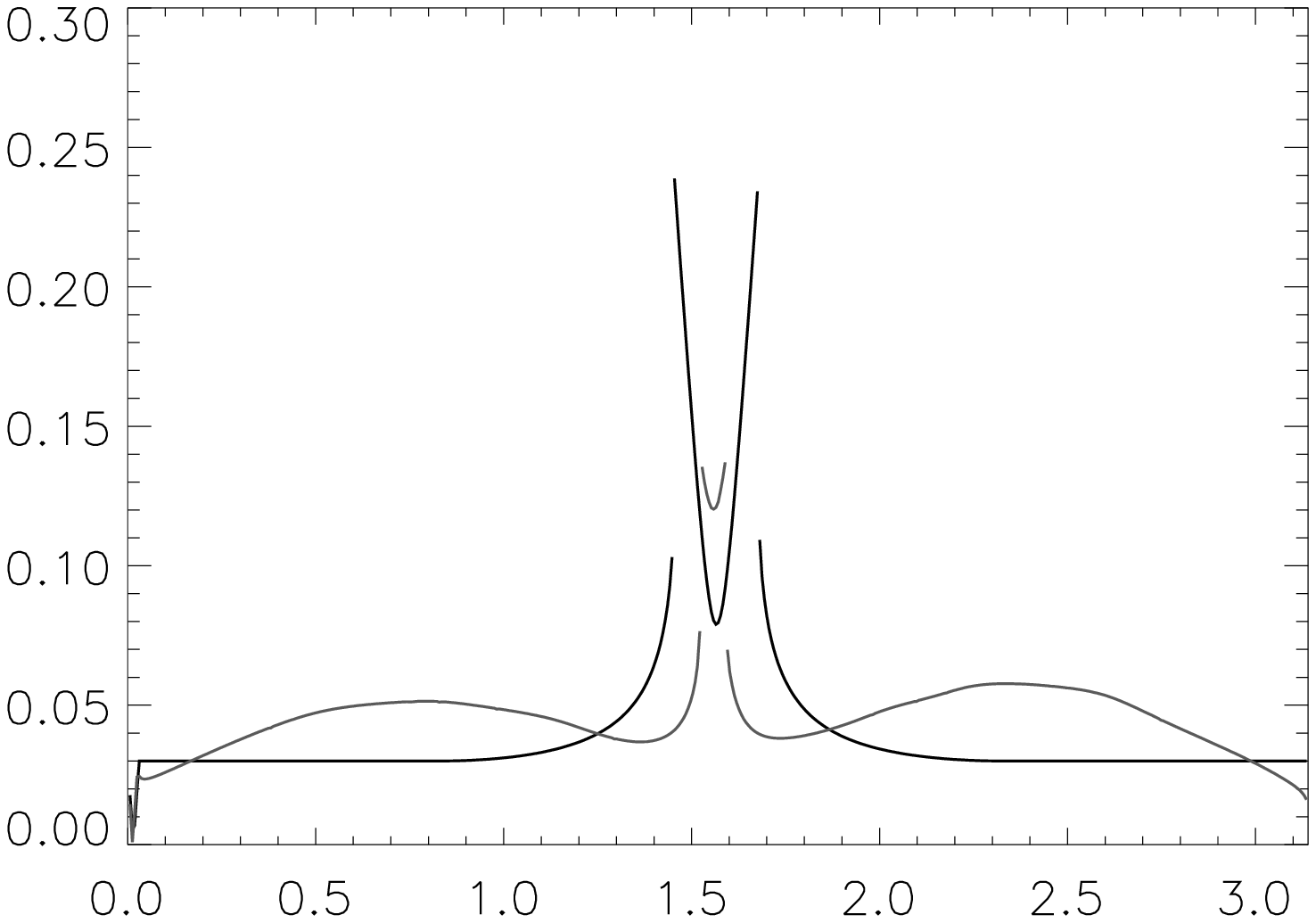}

\vspace*{-45mm}
\hspace*{-10mm}{\large $\Delta$\hspace*{85mm}$\Delta$}

\vspace*{37mm}
\hspace*{32mm}{\large $I_0$\hspace*{88mm}$I_0$}

\vspace*{1mm}
\hspace*{60mm}{\large (c)\hspace*{82mm}(d)}

\vspace*{3mm}
\noindent
\caption{The dependence of $\Delta(I_0,3\pi/2)$ on $I_0$ in the moonless system
(black line) and in the system with added exo-Moon (gray line).
The parameters are: $m_\rS=1$, $a_\rE=1$, $e_\rE=0$,
$m\rJ=0.05$, $a_\rJ=1.5$, $e_\rJ=0.1$, $\gamma_\rJ=\pi/16$, $\sigma=10$,
$\rho=0.5$ and $i=0$ (a), $i=\pi/64$ (b), $i=\pi/16$ (c) and $i=\pi/8$ (d).}
\label{fig4}\end{figure}

\begin{figure}[p]
\vspace*{-4mm}
\hspace*{-12mm}\includegraphics[width=8cm]{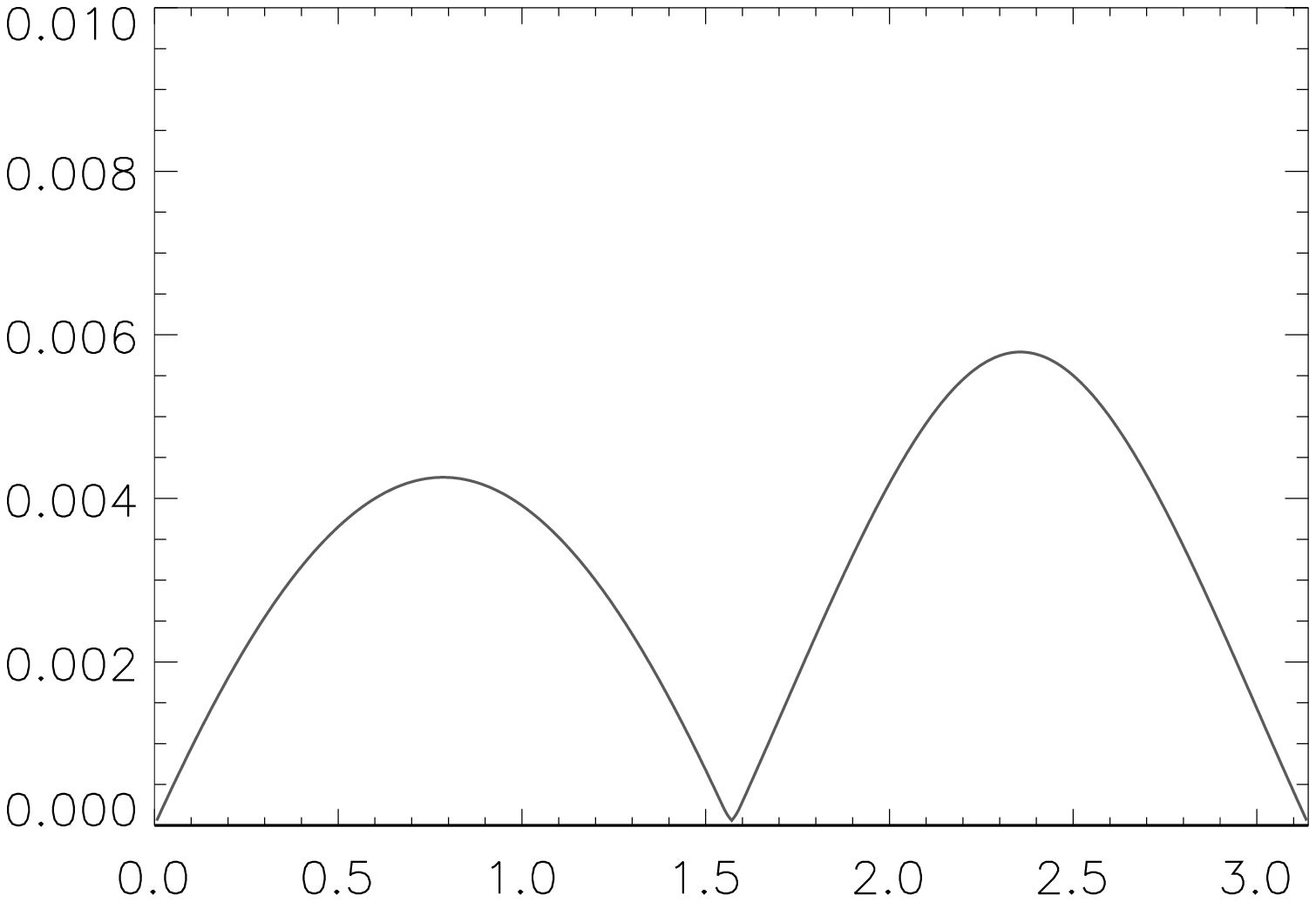}
\hspace*{10mm}\includegraphics[width=8cm]{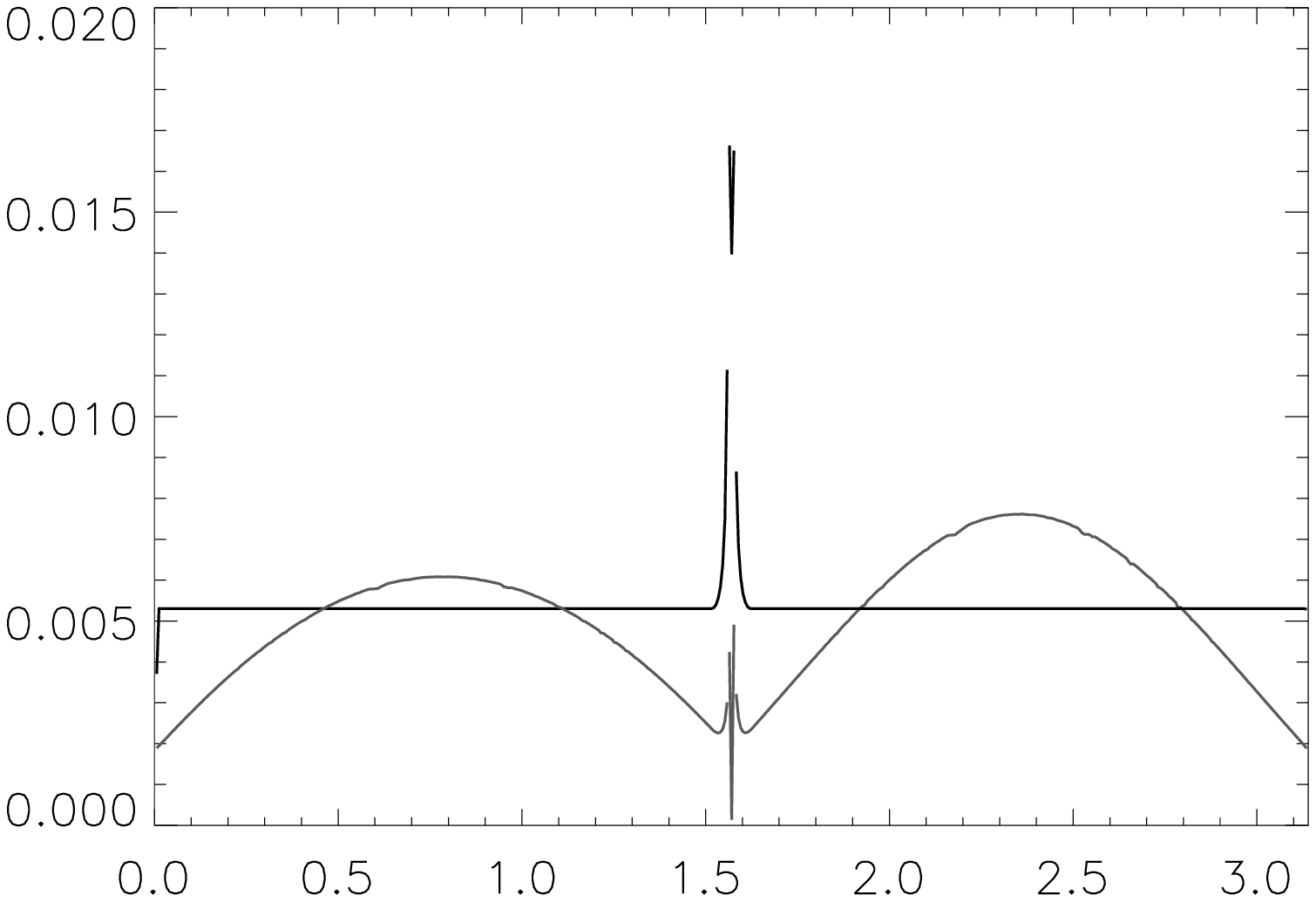}

\vspace*{-45mm}
\hspace*{-12mm}{\large $\Delta$\hspace*{87mm}$\Delta$}

\vspace*{37mm}
\hspace*{32mm}{\large $I_0$\hspace*{88mm}$I_0$}

\vspace*{1mm}
\hspace*{60mm}{\large (a)\hspace*{82mm}(b)}

\vspace*{4mm}
\hspace*{-12mm}\includegraphics[width=8cm]{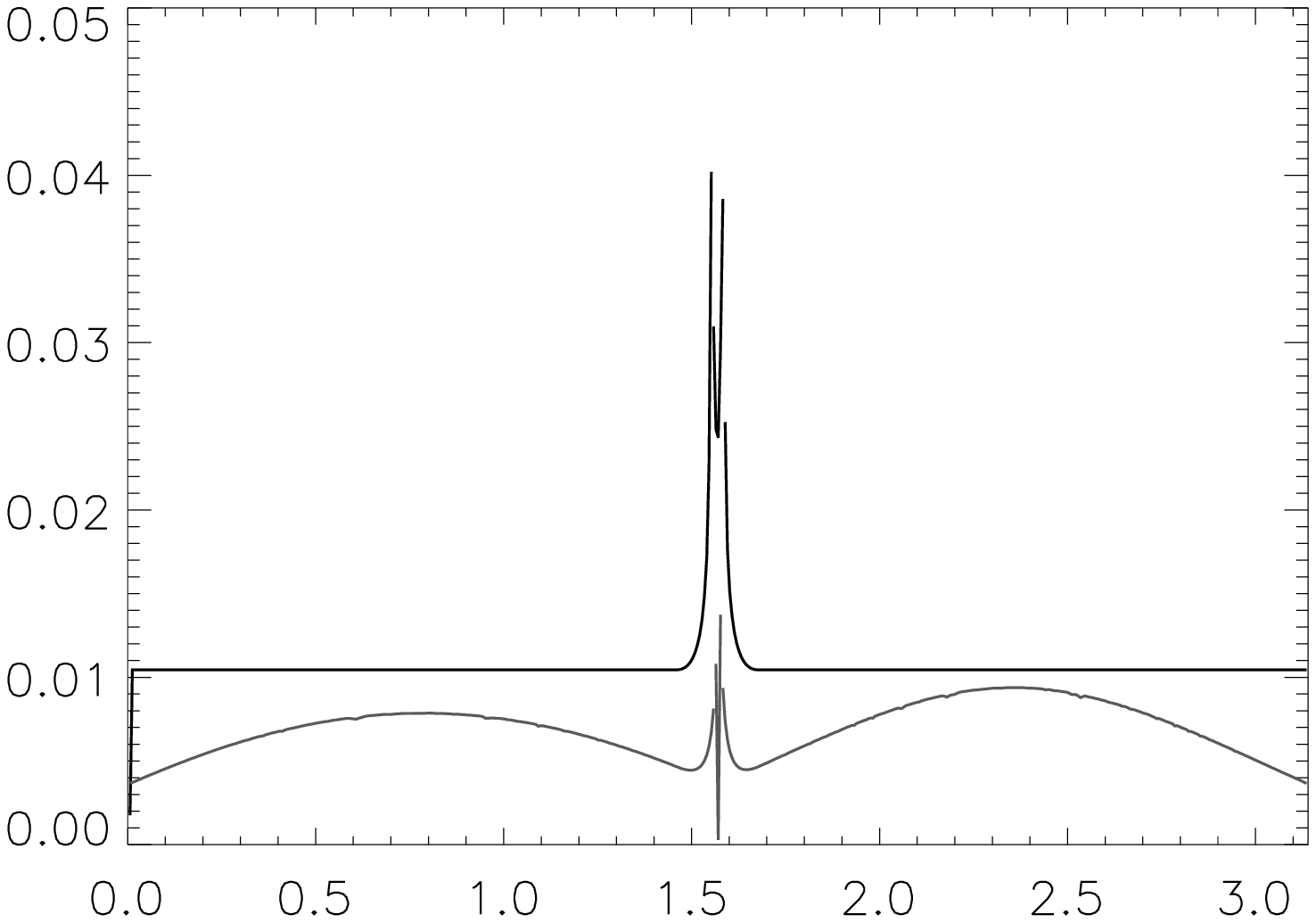}
\hspace*{10mm}\includegraphics[width=8cm]{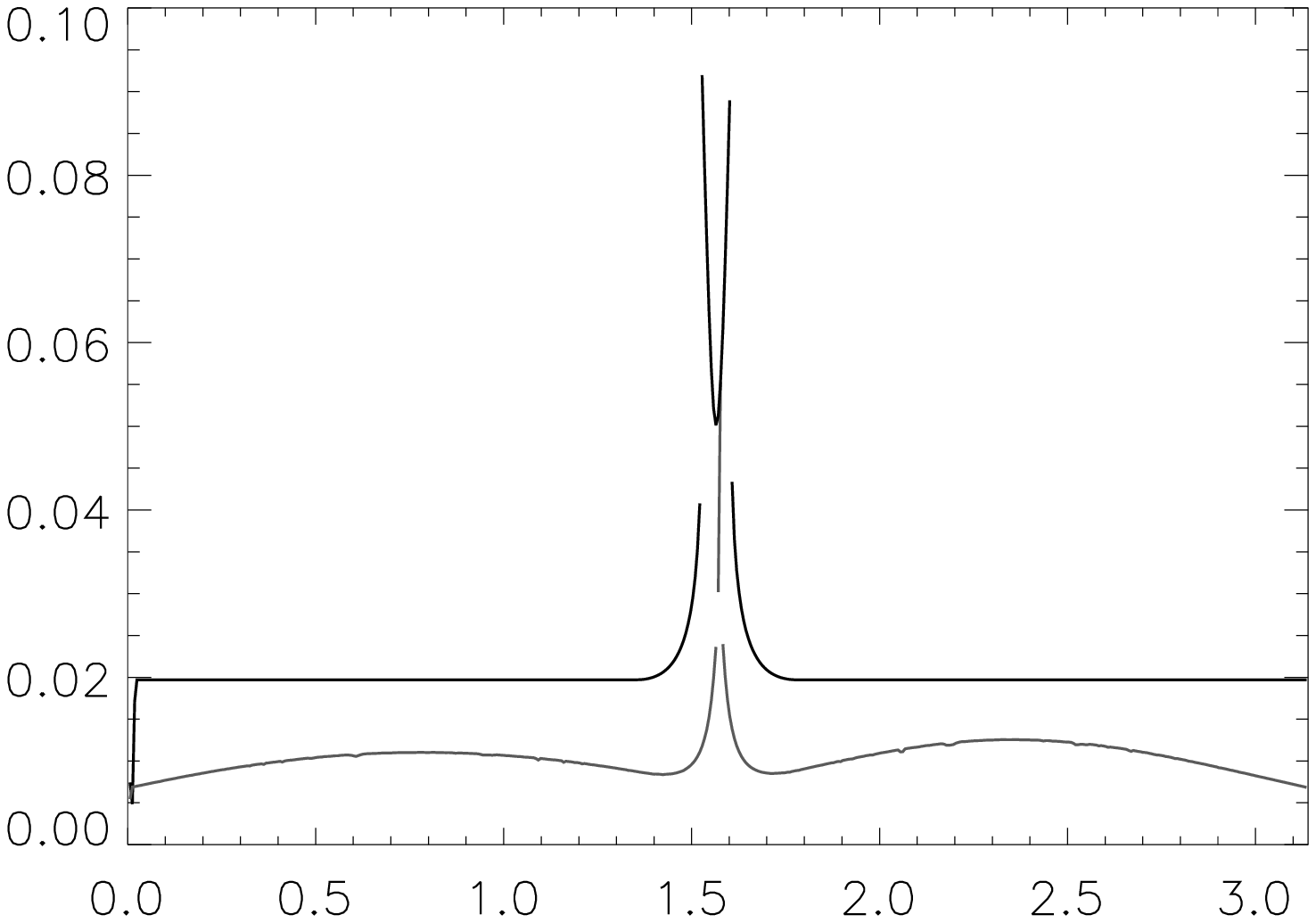}

\vspace*{-45mm}
\hspace*{-12mm}{\large $\Delta$\hspace*{87mm}$\Delta$}

\vspace*{37mm}
\hspace*{32mm}{\large $I_0$\hspace*{88mm}$I_0$}

\vspace*{1mm}
\hspace*{60mm}{\large (c)\hspace*{82mm}(d)}

\vspace*{3mm}
\noindent
\caption{The dependence of $\Delta(I_0,3\pi/2)$ on $I_0$ in the moonless system
(black line) and in the system with added exo-Moon (gray line).
The parameters are: $m_\rS=1$, $a_\rE=1$, $e_\rE=0$,
$m\rJ=0.05$, $a_\rJ=1.5$, $\sigma=10$, $\rho=1$ and $i=\pi/128$ and
$e_\rJ=0$, $\gamma_\rJ=0$ (a),
$e_\rJ=0.005$, $\gamma_\rJ=\pi/128$ (b),
$e_\rJ=0.01$, $\gamma_\rJ=\pi/64$ (c),
$e_\rJ=0.02$, $\gamma_\rJ=\pi/32$ (d).
}
\label{fig5}\end{figure}

We perform two series of computations. In the first one we start from an example
of subsection \ref{sec61} where the angle $i$ between the exo-Moon's orbit
and the ecliptic vanishes, implying that the impact is stabilising except
for $I_0$ near $\pi/2$. This is confirmed by numerical simulations
shown in Fig.~\ref{fig6}a. For small values of $i$ the impact is expected to continue
to be stabilising, the expectations are confirmed by numerical simulations
of Fig.~\ref{fig4}b.  As the angle is increased, for $I_0$ near $\pi/4$ and $3\pi/4$
the difference $\Delta^{P+M}-\Delta^{P}$ becomes positive and the difference
growths with $i$, see Fig.~\ref{fig4}c,d.

The second one starts from the system considered in subsection \ref{sec62}
with planets' orbits being circular and coinciding orbital planes.
Since $\Delta^{P}(I_0,h_0)=0$ for any initial condition in the moonless system,
the impact of the exo-Moon is destabilising (see fig.~\ref{fig5}a).
As the essentricity and inclination of the exo-Jupiter's orbit are increased,
the impact changes to stabilising, as shown in figs.~\ref{fig5}a-d.
Note that the impact is more destabilising around $I_0=\pi/4$ or $3\pi/4$ then
at the poles or near the equator. Such dependence of $\Delta^{P+M}-\Delta^{P}$
on $I_0$ might be a pecularity of two considered systems, or it may be of a
general type.

\section{Conclusion}\label{sec7}

In this paper we have studied the impact of a satellite on the evolution
of obliquity of a hypothetical exoplanet (an exo-Earth) at large times in
the case, where the orbital motions of celestial bodies affecting
the rotation of the exoplanet are quasiperiodic, the relevant frequencies
are not resonant and the orbit of the satellite is a Keplerian
ellipse which belong to a plane that keeps a constant angle with the
ecliptic while precessing with a prescribed angular velocity $\sigma$. Except
for the precession frequency, all other frequencies of the motion of
celestial bodies are order one.
The exo-Earth is assumed to be rigid and axially symmetric.

We follow the approach of \cite{pk20} where the evolution of obliquity
of a planet in a system comprised of stars and planets was studied by applying
time averaging over several fast variables with
non-resonant respective frequencies. At large times the evolution is
governed by a Hamiltonian involving six parameters which can be calculated
for prescribed masses and orbits of the celestial bodies. Without the satellite
the parameters are constants, while when the satellite is added they become
periodic in time with $\sigma$ being the respective frequency.

In a moonless system the Hamiltonian equations for the evolution of obliquity
are integrable \cite{pk20}.
Using the fact that the exo-Sun is substantially
heavier that any of other celestial bodies we derive approximations
for the range of obliquity as functions of initial conditions and
the six parameters involved in the Hamiltonian.

The full system, in general case, can not be integrated and the range of
obliquity should be found numerically. However, in some special cases
it can be proven analytically that the influence of exo-Moon is
stabilising or destabilising. Namely, it is stabilising if
orbital plane of the exo-Moon coincides with the ecliptic. It is also
stabilising if futher averaging over the fast variable $\sigma t$
is performed.
The influence of the exo-Moon is destabilising if orbits of all planets are
circular and their orbital planes coincide.

In this paper we have considered only direct influence of exo-Moon,
namely the torque from exo-Moon that effects the rotation of exo-Earth.
The indirect influence, caused by changing of the torque from exo-Sun
due to the modification of the orbit of exo-Earth by added exo-Moon can be
investigated by a similar approach.

Of course, it is highly interesting to consider planetary systems with
resonances: on the one hand, the presence of resonances drastically changes
the behaviour of an averaged system \cite{ar06,hl83} and, on the other hand,
resonances are abundant in the Solar system \cite{mu99} and therefore
we expect them to occur in other planetary systems as well.
One can conjecture that in the case when some of the
exo-Moon frequencies, $\sigma_n$, $\sigma_a$ or $\omega_2$,
are in resonance with some frequencies of the planetary motions the
impact of the exo-Moon is destabilising, because
averaging over resonant frequencies introduces additional slowly changing
variables, one for a resonance. The extra dimension(s) of the phase space
may lead to chaotic behaviour of the trajectories resulting on the increase of
the range of obliquity.

\subsection*{Acknowledgements}

Our research was partially financed by the grant 18-01-00820 from
the Russian foundation for basic research.

\appendix

\section{The range of nutation angle in a planetary system, comprised of
exo-Sun, exo-Earth and exo-Moon.}
\label{sec4}

In this section we approximate $\Delta$ in a system comprised of exo-Sun,
exo-Earth and exo-Moon under the assumption that the inclination of the
exo-Moon's orbital plane to the ecliptic is small following the approach of
section \ref{sec5}. The evolution of $\bf L$ in such a planetary system was
studied in a number of papers, see e.g., \cite{be75,smart,ti}.
The equations describing the motion of $\bf L$ on the celestial sphere
may have two, four or six steady states. The approximations for $\Delta$ that
we obtain are different depending on the number of the steady states,
they involve quantities which are functions of parameters of celestial bodies
and their orbital elements.

The coefficients $D_j=D_j^{(1)}+D^{(2)}_j$ that enter the averaged equations
on motion (\ref{gamav}),(\ref{aveG2})
are the sums of $D_j^{(1)}$ (\ref{coE}) that result from the torque from the Sun
and the ones $D^{(2)}_j$ (\ref{eq32}) from the Moon.
The mean Hamiltonian therefore is
\begin{equation}\label{aveG5}
\renewcommand{\arraystretch}{1.5}
\begin{array}{l}
\overline\cH_1={\displaystyle-{3\varepsilon\over2}(C_1-A_1)[
\sin^2I(-(\tilde D+D_1^{(2)})\sin^2h-(\tilde D+D_2^{(2)})\cos^2h}\\
+D_3^{(2)}+D_4^{(2)}\sin(2h))-\sin2I(D_5^{(2)}\sin h-D_6^{(2)}\cos h)].
\end{array}
\end{equation}
Substitution of (\ref{eq32}) and (\ref{eq33}) into (\ref{aveG5}) followed by
a series of algebraic transformations yield
\begin{equation}\label{aveG6}
\renewcommand{\arraystretch}{1.5}
\begin{array}{l}
\overline\cH_1=
{\displaystyle-{3\varepsilon\over2}(C_1-A_1)[
\sin^2I(-\tilde D+\Xi({1\over2}\sin^2i-\cos^2i)+}\\
{\Xi\over2}\sin^2i\cos(2\Omega-2h))
+{\Xi\over2}\sin2I\sin2i\cos(\Omega-h)].
\end{array}
\end{equation}

Using the generating function $F_2(h,H')=H'(h-\Omega_0-\sigma_a t)$, we obtain
that in the canonical coordinates $(H',h')=(H,h-\Omega_0-\sigma_a t)$
the Hamiltonian (\ref{aveG6}) takes the form
$$\cH'=\cH+{\partial F_2\over \partial t}=\cH-\sigma H'.$$
Therefore, (\ref{eqoH}), (\ref{eqFG}) imply that
\begin{equation}\label{gamav2}
{\rd h'\over\rd t}={\rho\over\sin I}
{\partial\cG'\over\partial I},\qquad
{\rd I\over\rd t}=-{\rho\over\sin I}
{\partial\cG'\over\partial h},
\end{equation}
where
\begin{equation}\label{ham2}
\renewcommand{\arraystretch}{1.5}
\begin{array}{l}{\displaystyle
\cG'(I,h')=\sin^2I(-\tilde D+\Xi({1\over2}\sin^2i-\cos^2i)+{\Xi\over2}\sin^2i
\cos2h')+}\\
{\displaystyle{\Xi\over2}\sin2I\sin2i\cos h'+{\sigma\over\rho}\cos I}
\end{array}
\end{equation}
and $\rho=3\varepsilon(C_1-A_1)/2G$.
Hence, in the new variables $(H,h')$ we have that
$\cG'=const$ along the trajectories. Note that the equation (\ref{ham2})
is invariant under the symmetry $h'\to-h'$
\begin{equation}\label{symm}
\cG'(I,h')=\cG'(I,-h').
\end{equation}
The equation is also invariant after the transformation
$(\sigma,I,h')\to(-\sigma,\pi-I,h'+\pi)$, therefore without the loss of
generality we consider non-negative $\sigma$'s only.

Equations (\ref{gamav2}) and (\ref{ham2}) can be re-written as follows
\begin{equation}\label{deri}
\renewcommand{\arraystretch}{2.2}
\begin{array}{l}
\displaystyle{{\rd h'\over \rd t}}={1\over\sin I}
(\sin2I(-D+\alpha\cos2h')+2\beta\cos2I\cos h'-\sigma\sin I)\\
\displaystyle{{\rd I\over\rd t}}=2\sin I\alpha\sin2h'+2\cos I\beta\sin h'.
\end{array}
\end{equation}
where
\begin{equation}\label{cons1}
\renewcommand{\arraystretch}{2.2}
\begin{array}{l}
D=\rho(\tilde D+\Xi(1-{3\over 2}\sin^2i/2)),\
\alpha=\rho\Xi\sin^2i/2,\\
\beta=\rho\Xi\sin2i/2\hbox{ and }\sigma\equiv\sigma_a.
\end{array}
\end{equation}

\begin{remark}
Due to the assumption that the angle $i$ is small, the coefficients of
(\ref{deri}) satisfy $\alpha\ll\beta$. To compare $D$ with $\beta$
we note that $D\sim \rho fm_S/R_E^3$, while
$\beta=\rho\Xi\sin2i/2\sim \rho\sin ifm_E/R_M^3$.
Hence, depending on a planetary
systems they may be comparable, or one can be much larger then the other. In
particular, for the Earth and the Moon $D/\Xi\sim m_SR_M^3/m_ER_E^3\approx 30$,
i.e. $D$ is significantly larger than $\beta$. To compare $D$ with $\sigma_a$
we recall that $D\sim (C-A)fm_S/GR_E^3$ while
$\sigma_a\sim\omega_E^2/\omega_M$, $\omega_E^2=fm_S/R_E^3$ and
$\omega_M^2=fm_E/R_M^3$.
Therefore $D/\sigma_a\sim (C-A)G^{-1}f^{1/2}m_E^{1/2}R_M^{-3/2}$. For the
Earth the ratio is very small (about $10^{-10}$), while this may not be the
case for other planetary systems.
\end{remark}

Steady states of (\ref{deri}) satisfy
$${\partial \cG'\over \partial h'}={\partial \cG'\over \partial I}=0,$$
which can be re-written as
\begin{equation}\label{ss0}
\renewcommand{\arraystretch}{1.5}
\begin{array}{l}
\sin2I(-D+\alpha\cos2h')+2\beta\cos2I\cos h'-\sigma'\sin I=0;\\
\sin h'(\sin I\alpha\cos h'+\cos I\beta)=0.
\end{array}\end{equation}
Hence, the steady states can be found from the following equations
\begin{equation}\label{ss1}
h'=0,\pi,\quad \sin2I(-D+\alpha)\pm2\beta\cos2I-\sigma'\sin I=0
\end{equation}
\begin{equation}\label{ss2}
\cos h'=\displaystyle{{-\cos I\beta\over\sin I\alpha}},\quad
\sin2I(-D+\alpha\cos2h')+2\beta\cos2I\cos h'-\sigma\sin I=0.
\end{equation}

Below we assume that similarly to the Earth $D\gg\beta$. Under this assumption
the steady states (\ref{ss1}) are:
\begin{equation}\label{ss11}
h'_{ss,1}=0,\ I_{ss,1}\approx{2\beta\over D-\alpha+\sigma};\quad
h'_{ss,2}=0,\ I_{ss,2}\approx\pi-{2\beta\over D-\alpha+\sigma};
\end{equation}
\begin{equation}\label{ss12}
h'_{ss,3}=0,\ \cos I_{ss,3}\approx{-\sigma\over 2(D-\alpha)},\quad
h'_{ss,4}=\pi,\ \cos I_{ss,4}\approx{-\sigma\over 2(D-\alpha)},
\end{equation}
where the latter two
\footnote{
To find the steady states (\ref{ss12}) we re-write the second equation in
(\ref{ss1}) as
$$\sin I(2\cos I(-D+\alpha)-\sigma)\pm2\beta\cos2I=0,$$
which due to the assumption $D\gg\beta$ implies that unless $\sin I$ is
small (this gives the steady states (\ref{ss11})\,), the solution to the
equation satisfies $|2\cos I(-D+\alpha)-\sigma|\ll\beta$.
}
exist only if $|\sigma|\le D-\alpha$.

To find steady states satisfying (\ref{ss2}), we note that $\beta\gg\alpha$
implies that $\cos I\ll\sin I$, i.e. that $I\approx\pi/2$. Hence, we can write
that $\cos h'\approx-\beta\cos I/\alpha$. Substituting this into the
second equation (\ref{ss2}) we obtain that
$$\cos I_{ss,5}\approx{-\alpha\sigma\over 2\alpha(D+\alpha)+\beta^2},\quad
\cos h_{ss,5}\approx{\beta\sigma\over 2\alpha(D+\alpha)+\beta^2}.$$
(And $I_{ss,6}=I_{ss,5}$, $h'_{ss,6}=-h'_{ss,5}$.)
Therefore, these steady states exist whenever
$${\beta|\sigma|\over 2\alpha(D+\alpha)+\beta^2}\le 1.$$

\begin{figure}[h]
{\large
\vspace*{1mm}
\hspace*{-15mm}\includegraphics[width=7cm]{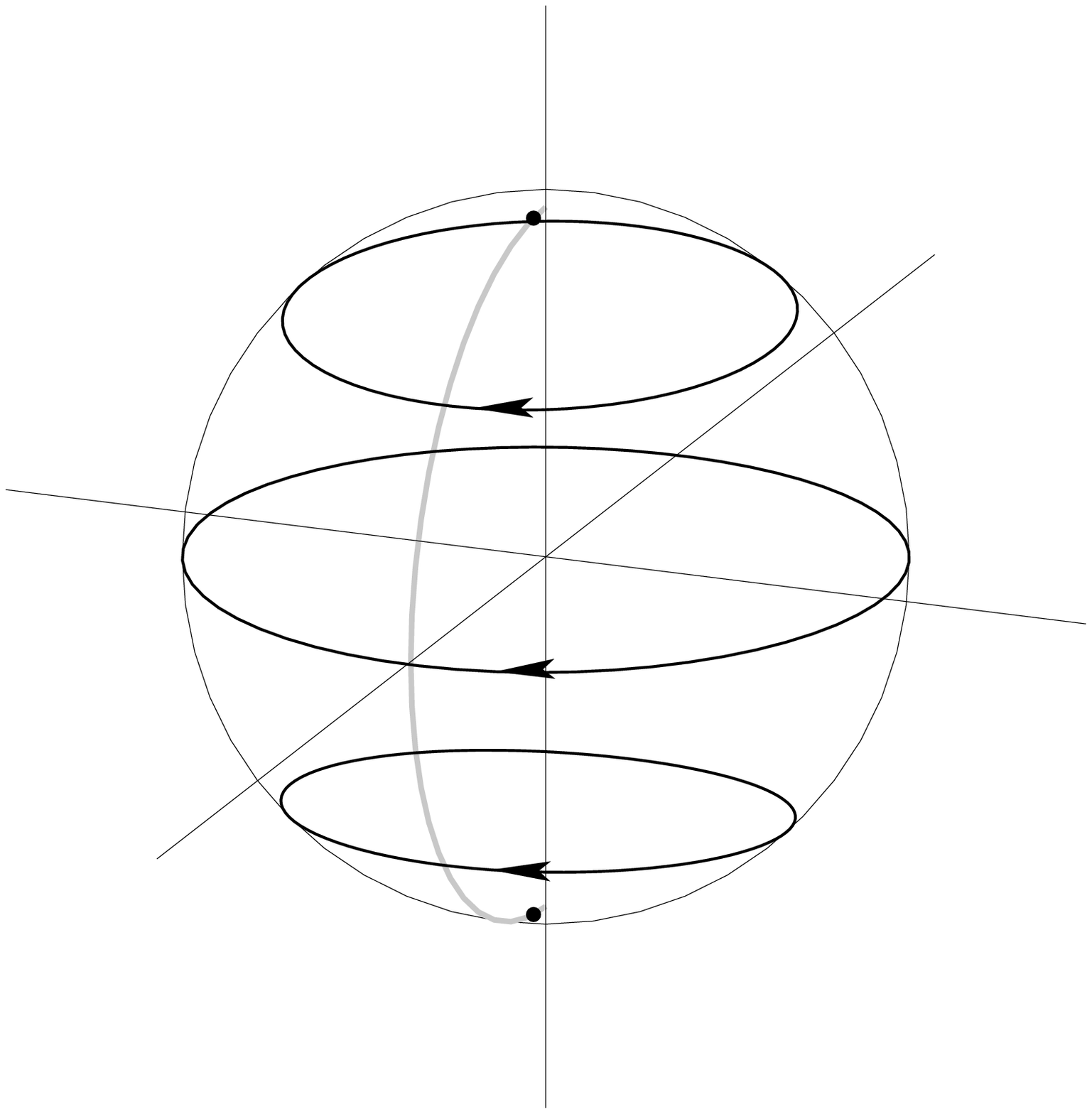}\hspace*{-15mm}
\includegraphics[width=7cm]{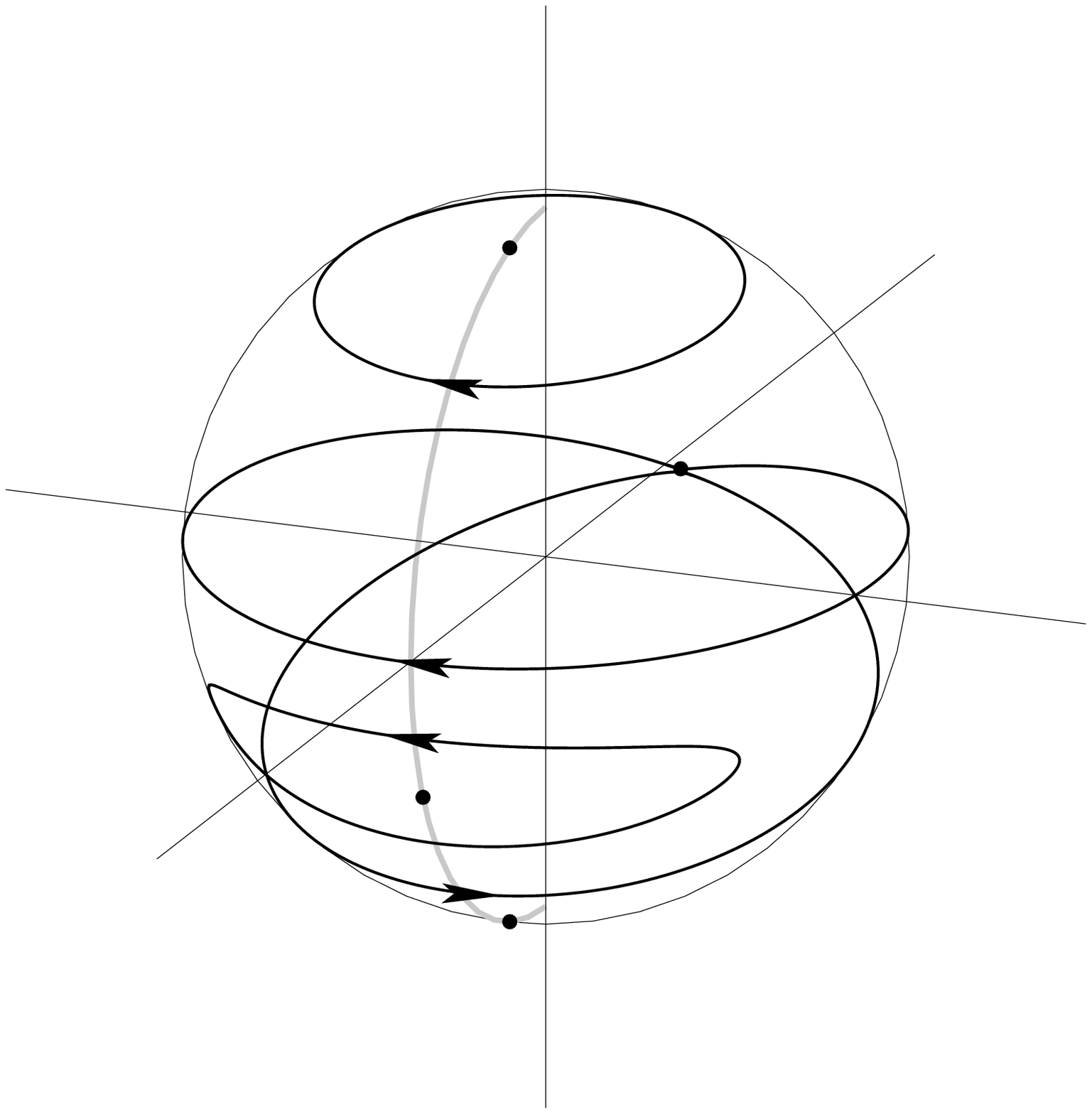}\hspace*{-15mm}
\includegraphics[width=7cm]{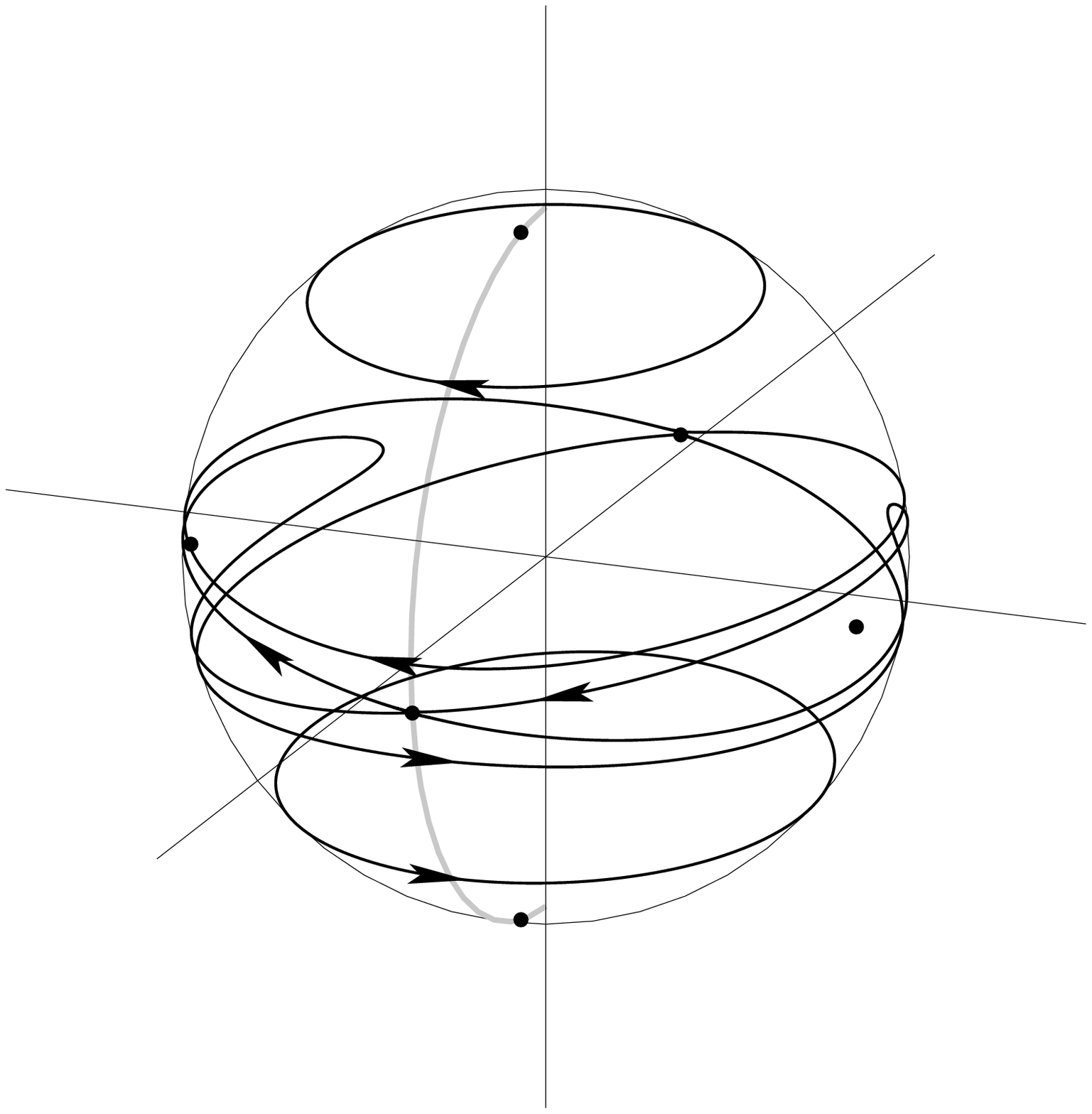}
\vspace*{4mm}

\vspace*{-33mm}
\hspace*{4mm}$\xi$\hspace*{54mm}$\xi$\hspace*{54mm}$\xi$

\vspace*{-18mm}
\hspace*{47mm}$\eta$\hspace*{54mm}$\eta$\hspace*{54mm}$\eta$

\vspace*{-40mm}
\hspace*{17mm}$\zeta$\hspace*{54mm}$\zeta$\hspace*{54mm}$\zeta$

\vspace*{52mm}
\hspace*{40mm}(a)\hspace*{54mm}(b)\hspace*{54mm}(c)

}
\noindent
\caption{
Motion of $\bL$ (\ref{comL}) on the celestial sphere
for $I$ and $h$ solving
equations (\ref{gamav}), (\ref{aveG2}) in cases I,
$D=1$, $\alpha=0.05$, $\beta=0.5$, $\sigma=10$ (a), II
$D=1$, $\alpha=0.02$, $\beta=0.2$, $\sigma=0.5$ (b) and
$D=1$, $\alpha=0.02$, $\beta=0.1$, $\sigma=0.1$ III (c).}
\label{fig6}\end{figure}

Overall, the system may have two, four or six steady states. The above
conditions for the existence of the steady states can be summarised as follows:
\begin{equation}\label{steadystates}
\renewcommand{\arraystretch}{1.5}
\begin{array}{l|l|l}
\hbox{case I} &\sigma> D-\alpha & \hbox{two steady states (\ref{ss11}), both centers}\\
\hline
\hbox{case II} &\sigma< D-\alpha \hbox{ and}& \hbox{four steady states: two centers
(\ref{ss11}),}\\
&\beta\sigma> 2\alpha(D+\alpha)+\beta^2&
\hbox{one center and one saddle (\ref{ss12})}\\
\hline
\hbox{case III}&\beta\sigma< 2\alpha(D+\alpha)+\beta^2&
\hbox{six steady states: two centers (\ref{ss11}),}\\
&&\hbox{two saddles (\ref{ss12}), two centers (\ref{ss2})}.
\end{array}\end{equation}
Evolution of $\bL$ on the celestial sphere in these three cases is shown in
Fig.~\ref{fig6}

Maxima and minima of $I$ for a particular trajectory $(I(t),h'(t))$ are achieved
at $\rd I/\rd h'=0$, which due to (\ref{deri}) takes place at
$$h'=0,\pi\hbox{ or at the points where }
\cos h'={-\cos I\beta\over\sin I\alpha}.$$
Hence, for a trajectory through $(I_0,h_0)$ the extreme values of $I$, which
we label by $I^{(0)}$, $I^{(\pi)}$ and $I^{(*)}$ achieved at the points
given above, can be found by solving the equations
\begin{equation}\label{extrema}
\renewcommand{\arraystretch}{1.5}
\begin{array}{l}
\cG'(I^{(0)},0)=s;\ \cG'(I^{(\pi)},\pi)=s;\\
\cG'(I^{(*)},h^{(*)})=s,\quad
\cos h^{(*)}=\displaystyle{-\cos I^{(*)}\beta\over\sin I^{(*)}\alpha};
\end{array}\end{equation}
for $s=\cG'(I_0,h_0)$. The last equation in (\ref{extrema}) can not be solved
for all trajectories. If it can be solved, it has two solutions
$(I^{(*)},h^{(*)})$ and $(I^{(*)},-h^{(*)})$, due to (\ref{symm}).

Below we study how the range of nutation angle $\Delta(I_0,h_0)$ (\ref{Delta})
for a trajectory $(I(t),h'(t))$ with $(I(0),h'(0))=(I_0,0)$ depends on $I_0$,
considering individually each of three cases outlined in (\ref{steadystates}).
Recall that we assume $D\gg\beta$ which implies that for a particular
trajectory the difference $I(t)-I_0$ is small and we can write
$I(t)=I_0+I_1(t)$.
In case I when no heteroclinic equilibria exist the function $I(t)$
can be regarded as a function of $h'$. Writing
\begin{equation}\label{appro}
\renewcommand{\arraystretch}{1.5}
\begin{array}{l}
s=\cG'(I_0,\pi)=\sin^2I_0(-D+\alpha)-\beta\sin2I_0+\sigma\cos I_0,\\
\sin^2I\approx\sin^2 I_0+I_1(h')\sin2I_0+I_1^2(h')\cos2I_0,\\
\sin2I\approx\sin2I_0+I_1(h')2\cos2I_0-I_1^2(h')2\sin2I_0,\\
\cos I\approx\cos I_0-I_1(h')\sin I_0I_1(h')-I_1^2(h')\cos I_0/2
\end{array}\end{equation}
and substituting these into
\begin{equation}\label{ham2n}
\cG'(I,h')\equiv\sin^2I(-D+\alpha\cos2h')+\beta\sin2I\cos h'+\sigma\cos I=s
\end{equation}
we obtain that $I_1^{(\pi)}$ can be found from
\begin{equation}\label{I01}
a(I_1^{(\pi)})^2+bI_1^{(\pi)}+c=0,
\end{equation}
where
\begin{equation}\label{I02}
\renewcommand{\arraystretch}{1.5}
\begin{array}{l}
a=-D\cos2I_0+2\beta\sin2I_0-{1\over2}\sigma\cos I_0\\
b=-D\sin2I_0-2\beta\cos2I_0-\sigma\sin I_0\\
c=-2\beta\sin2I_0.
\end{array}\end{equation}
(In the first two lines in (\ref{I02}) we omit $\alpha$ using the fact
that $D'\gg\alpha$.)

A trajectory $(I(t),h'(t))$ has two extrema of $I(t)$, which as discussed above
are achieved at $h'=0$ and $h'=\pi$. Hence, we have that
$\Delta(I_0,0)=|I_1^{(\pi)}|$. The formula for the roots of cubic equation
therefore implies
\begin{equation}\label{case11}
\Delta(I_0,0)==\biggl|{|b|-(b^2-4ac)^{1/2}\over 2a}\biggr|
\end{equation}
with $a,b$ and $c$ given in (\ref{I02}).

Two additional extrema (see (\ref{extrema})\,) are achieved at $h^{(*)}$,
where\\
$\cos h^{(*)}=-\cos I^{(*)}\beta/\sin I^{(*)}\alpha$. Therefore,
they exist only for trajectories such that\\
$|\cos I_0\beta/\sin I_0\alpha|\le 1$.
(Here we use the fact that $I^{(*)}\approx I_0$.) Since
$\alpha\ll\beta$, this implies $I_0\approx\pi/2$ and
$\cos h^{(*)}\approx-\cos I_0\beta/\alpha$. From (\ref{I01}) and (\ref{I02}),
for such trajectories the expression for $I_1^{(\pi)}$ simplifies to
\begin{equation}\label{case12}
I_1^{(\pi)}={4\beta\cos I_0\over\sigma}.
\end{equation}
By the same algebraic transformations as above we obtain that
\begin{equation}\label{case13}
I_1^{(*)}={2\alpha-\beta\cos I_0\over\sigma}.
\end{equation}
The range of nutation angle is the maximum of $|I_1^{(\pi)}|$ and
$|I_1^{(\pi)}-I_1^{(*)}|$,
therefore (\ref{case12}) and (\ref{case13}) imply that
\begin{equation}\label{caseIf}
\Delta(I_0,0)={2\alpha+2\beta|\cos I_0|\over\sigma}.
\end{equation}

In case II heteroclinic trajectories through the steady state $(I_{ss,4},\pi)$
split the celestial sphere into three regions, comprised of a center and
a set of trajectories around this steady state.
Two of the centers are located near poles and we call the respective regions
{\it polar}, while the remaining one
\footnote{
To show that the steady state $(I_{ss,3},0)$ is a center we note that
in the coordinates $(I_1=I(t)-I_{ss,3},h_1=h'(t)-h_{ss,3})$ nearby
trajectories satisfy the equation
$I_1^2(D+2\beta\sin2I_{ss,3})+h_1^2(-2\alpha\sin^2I_{ss,3}-
\beta\sin2I_{ss,3}/2)=s$. In case II we have that
$(D+2\beta\sin2I_{ss,3})(-2\alpha\sin^2I_{ss,3}-\beta\sin2I_{ss,3}/2)>0$,
which implies the statement.}
$(I_{ss,3},0)$, in general, is not.
By contrast, we call {\it equatorial} the region near $(I_{ss,3},0)$.
Inside a region the range of obliquity depends continuously on the
initial condition, while it is discontinuous when crossing a boundary.

For a trajectory $(I(t),h'(t))$ inside the equatorial region the maximal and
minimal value of $I(t)$, $I_{\max}$ and $I_{\min}$, are both achieved at $h'=0$.
Moreover, as we noted above $I(t)$ does not differ much from $I_0$ and
$(I_{ss,3},0)$ is a center, therefore $I_{\max}-I_{ss,3}\approx I_{ss,3}-I_{\min}$.
The initial $(I_0,0)$ corresponds either to the maximum or to the minimum
of $I(t)$, implying that
\begin{equation}\label{caseIIb}
\Delta(I_0,0)=I_{\max}-I_{\min}\approx 2|I_{ss,3}-I_0|.
\end{equation}
The region is bounded by a heteroclinic trajectory through $(I_{ss,4},\pi)$,
where $I_{ss,4}=I_{ss,3}$ (see (\ref{ss12})\,). The initial condition
$(I_0,0)$ belongs to this region if
$I_{ss,3}-\delta_{het}<I_0<I_{ss,3}+\delta_{het}$. Following the same ideas that
are used to calculate $\Delta$, we obtain that the value of
$\delta_{het}$ is a solution to the following equation
$$\cG'(I_{ss,3}+\delta_{het},0)=\cG'(I_{ss,4},\pi).$$
Solving the equation we find that
$$\delta_{het}=\biggl({2\beta|\sin2I_{ss,3}|\over D}\biggr)^{1/2}.$$

In polar regions the range is calculated similarly to case I.
Namely, when a trajectory through $(I_0,h_0)$ has one minimum and one maximum
then the range is given by (\ref{case11}),(\ref{I02}). In the case of
four extrema and $I_0\le\pi/2$ we have
\begin{equation}\label{case23}
\Delta=|I_1^{(*)}|={4\beta^2\cos^2 I_0+2\alpha^2-2\alpha\beta\cos I_0
\over-2D\alpha^2/beta+2\alpha\beta+\alpha\sigma}.
\end{equation}
If there are four extrema and $I_0\ge\pi/2$ then
\begin{equation}\label{case24}
\Delta=I_1^{(*)}-I_1^{(\pi)},
\end{equation}
where $I_1^{(*)}$ is given by (\ref{case23}) and
\begin{equation}\label{case25}
I_1^{(\pi)}=-{4\beta\cos I_0\over 2D\cos I_0+4\beta+\sigma}.
\end{equation}

In case III there are three steady states inside the equatorial region.
One is $(I_{ss,3},0)$ which now is a saddle and the other two,
$(I_{ss,5},h_{ss,5})$ and $(I_{ss,6},h_{ss,6})$, are centers. The meridian
of initial conditions $(I_0,0)$ does not cross the boundaries of the emerging
regions around the latter steady states.

For a trajectory inside the equatorial region that takes extreme values only at
$h'=0$ the range can be found from (\ref{caseIIb}). If additional
extrema at $h'=h_{*}$ (and also at $h'=-h_{*}$) exist then both
maximal and minimal values along a trajectory are taken at this value of $h'$.
The maximal and minimal values of $I_1(t)$ are solutions to the cubic equation
(\ref{I01}). They are
\begin{equation}\label{caseIIIc}
|I_1^{(\min,\max)}|={-b\pm(b^2-4ac)^{1/2}\over 2a},
\end{equation}
where
\begin{equation}\label{caseIIId}
a=D,\ b=\cos I_0(-2D+{2\beta^2\over\alpha})-\sigma,\
c=-2\alpha+2\beta\cos I_0.
\end{equation}
Therefore
\begin{equation}\label{caseIIIe}
\Delta(I_0,0)=I_1^{(\max)}-I_1^{(\min)}={|b|\over|a|}=
{|\cos I_0(-2D\alpha+2\beta^2)-\sigma\alpha|\over D\alpha}.
\end{equation}

The results can be summarised as follows
\begin{equation}\label{final}
\renewcommand{\arraystretch}{1.5}
\begin{array}{l|l|l}
\hbox{case I} & |I_0-\pi/2|>\alpha\beta & \hbox{expressions
(\ref{I02}) and (\ref{case11})}\\
              & |I_0-\pi/2|<\alpha\beta & \hbox{expression (\ref{caseIf})}\\
\hline
\hbox{case II} & |I_0-\pi/2|>\alpha\beta & \hbox{expressions
(\ref{I02}) and (\ref{case11})}\\
&\hbox{and} |I_0-I_{ss,3}|>\delta_{het}& \\
              & |I_0-\pi/2|<\alpha\beta & \hbox{expressions
(\ref{case23}) or (\ref{case24}),(\ref{case23}),(\ref{case25})}\\
	      & |I_0-I_{ss,3}|<\delta_{het}& \hbox{expression (\ref{caseIIb})}\\
\hline
\hbox{case III} & |I_0-I_{ss,3}|>\delta_{het}& \hbox{expressions
(\ref{I02}) and (\ref{case11})}\\
       & |I_0-I_{ss,3}|<\delta_{het}& \hbox{expression (\ref{caseIIb})}\\
&\hbox{and} |I_0-I_{ss,3}|>\alpha\beta& \\
              & |I_0-\pi/2|<\alpha\beta & \hbox{expressions (\ref{caseIIId})
and (\ref{caseIIIe})}
\end{array}\end{equation}

\begin{figure}[h]
\vspace*{1mm}
\hspace*{-12mm}\includegraphics[width=8cm]{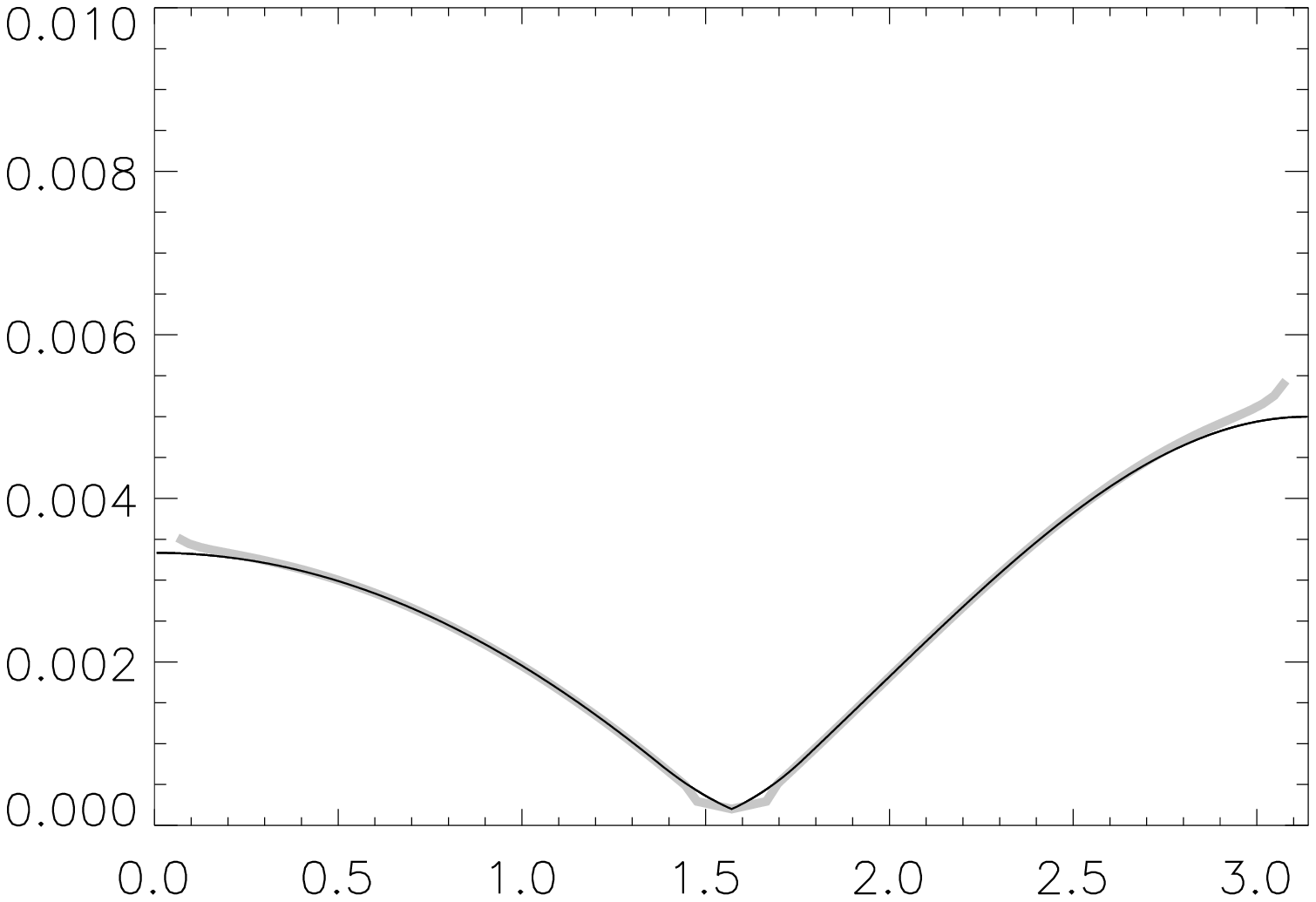}
\hspace*{10mm}\includegraphics[width=8cm]{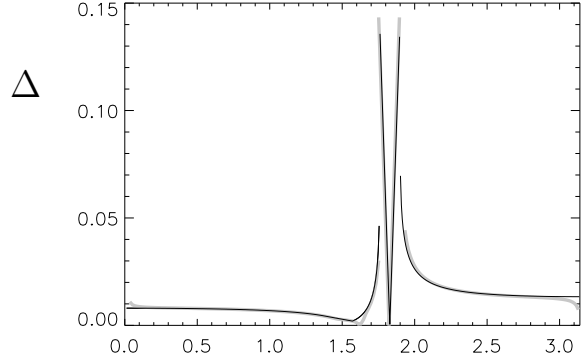}

\vspace*{-45mm}
\hspace*{-12mm}{\large $\Delta$\hspace*{87mm}$\Delta$}

\vspace*{37mm}
\hspace*{32mm}{\large $I_0$\hspace*{88mm}$I_0$}

\vspace*{1mm}
\hspace*{60mm}{\large (a)\hspace*{82mm}(b)}

\vspace*{3mm}
\noindent
\caption{The dependence of $\Delta(I_0,0)$ on $I_0$ found
by integrating in time equations (\ref{deri}) (black line) and from
approximations (\ref{final}) (gray line) in cases I
$D=1$, $\alpha=0.001$, $\beta=0.01$, $\sigma=10$ (a) and II
$D=1$, $\alpha=0.0005$, $\beta=0.005$, $\sigma=0.5$ (b).}
\label{fig7}\end{figure}

\end{document}